\pdfoutput=1

\documentclass[a4paper,11pt]{article}
\usepackage{slashed}
\usepackage[affil-sl,auth-sc]{authblk}
\usepackage{amssymb,amsmath,amsbsy}
\usepackage{mathrsfs}
\usepackage{mathbbol}
\usepackage{hyperref}
\usepackage{pinlabel}

\usepackage[top=1.2in, bottom=1.2in,left=0.9in,includefoot]{geometry}     
\usepackage{graphicx}
\usepackage{cite}
\usepackage{float}
\usepackage{caption}
\usepackage{subfig}

\def\code{{\tt SUSY\_FLAVOUR}}

\def\theequation{\arabic{section}.\arabic{equation}}
\def\eq#1{eq.~(\ref{#1})}

\def\eqs#1#2{eqs.~(\ref{#1}) and (\ref{#2})}

\def\eqst#1#2{eqs.~(\ref{#1})--(\ref{#2})}

\def\Ref#1{ref.~\cite{#1}}

\def\Refs#1{refs.~\cite{#1}}

\newcommand{\Br}{\mathcal{B}}
\newcommand{\tqh}{t\to q\, h}
\newcommand{\tuh}{t\to u\, h}
\newcommand{\tch}{t\to c\, h}
\newcommand{\ie}{{\it i.e., }}

\usepackage{color}

\definecolor{orange}{rgb}{0.9,0.2,0}
\definecolor{brown}{rgb}{0.7,0.3,0.2}
\definecolor{fuxia}{rgb}{1,0,1}
\definecolor{skyblue}{rgb}{0,0.1,0.9}
\definecolor{violetred}{rgb}{0.8,0.13,0.56}
\definecolor{deeppink}{rgb}{1.00,0.08,0.5}
\definecolor{pink}{rgb}{1.00,0.75,0.80}
\definecolor{orchid}{rgb}{0.85,0.44,0.84}
\definecolor{lightpink}{rgb}{1.00,0.71,0.76}
\definecolor{bluish}{rgb}{0,0.6,0.8}
\definecolor{greyblue}{rgb}{0.3,0.3,0.7}

\title{\bf Rare Top-quark Decays to Higgs boson in  MSSM}

\author{A. Dedes$^{1,2}$\footnote{email: {\tt adedes@cc.uoi.gr}},~ 
M. Paraskevas$^{1}$\footnote{email: {\tt mparask@grads.uoi.gr}},~
J. Rosiek$^{3}$\footnote{email: {\tt janusz.rosiek@fuw.edu.pl }},~
K. Suxho$^{1}$\footnote{email: {\tt csoutzio@cc.uoi.gr}},~
K. Tamvakis$^{1}$\footnote{email: {\tt tamvakis@uoi.gr}}}
\affil{\small $^{1}$Department of Physics, Division of Theoretical Physics, \\
 University of Ioannina, GR 45110, Greece}
\affil{\small $^{2}$University of Athens, Physics Department, \\
Nuclear and Particle Physics Section, GR 15771 Athens, Greece}
\affil{\small $^{3}$Institute of Theoretical Physics, Warsaw University, \\ 
Hoza 69, 00-681 Warsaw, Poland}

\date{November 14, 2014}                                          

\begin{document}

\maketitle

\begin{abstract}
In full one-loop generality and in next-to-leading order in QCD, we
study rare top to Higgs boson flavour changing decay processes $\tqh$
with $q=u,c$ quarks, in the general MSSM with R-parity conservation.
Our primary goal is to search for enhanced effects on $\Br(\tqh)$ that
could be visible at current and high luminosity LHC running.
To this end, we perform an analytical expansion of the amplitude in
terms of flavour changing squark mass insertions that treats both
cases of hierarchical and degenerate squark masses in a unified way.
We identify two enhanced effects allowed by various constraints: one
from holomorphic trilinear soft SUSY breaking terms and/or right
handed up squark mass insertions and another from non-holomorphic
trilinear soft SUSY breaking terms and light Higgs boson masses.
Interestingly, even with $\mathcal{O}(1)$ flavour violating effects in
the, presently unconstrained, up-squark sector, SUSY effects on
$\Br(\tqh)$ come out to be unobservable at LHC mainly due to leading
order cancellations between penguin and self energy diagrams and the
constraints from charge- and colour-breaking minima (CCB) of the MSSM
vacuum.  An exception to this conclusion may be effects arising from
non-holomorphic soft SUSY breaking terms in the region where the
CP-odd Higgs mass is smaller than the top-quark mass but this scenario
is disfavoured by recent LHC searches.  Our calculations for $\tqh$
decay are made available in \code{} numerical library.

\end{abstract}


\newpage
\section{Introduction}
\setcounter{equation}{0}
\label{intro}

The last fundamental elementary particles discovered during the last
20 years are the tau-neutrino by DONUT
Collaboration~\cite{Kodama:2000mp}, the top quark at
Tevatron~\cite{Abachi:1995iq, Abe:1995hr} with mass $m_{t}=172.5$ GeV
and the Higgs boson~\cite{Englert:1964et,Higgs:1964pj,Guralnik:1964eu}
at LHC~\cite{Chatrchyan:2012ufa,Aad:2012tfa}, with mass $m_{h} \approx
126$ GeV.  Among them, the top quark has been and will be produced in
large numbers at LHC, allowing for increasingly accurate measurements
of its properties.  LHC operating at c.m.  energy of 7 and 8 TeV has
already collected about two-million $t\bar{t}$-pairs.  It is therefore
timely to examine the possibility of rare, flavour-changing (FC), top
decays to the light up-quarks, $u$ or $c$, and the Higgs boson~$h$,
\begin{eqnarray}
t \to u\, h \;,  \quad \mathrm{or} \quad t\to c\, h \;.\label{tch}
\end{eqnarray} 
We collectively denote these processes as $\tqh$ with $q=u,c$.  The
Higgs boson field $h$ is understood as one of the possible scalar
fields that couples to up-quarks and has mass smaller than that of the
top-quark.

If the decays $\tqh$ are governed only by the Standard Model
(SM)~\cite{Weinberg:1967tq} dynamics they would never be observed at
LHC because their branching ratios, $\Br(t\to u\, h)_{\rm SM} \approx
4\times 10^{-17}$ and $\Br(t\to c\, h)_{\rm SM} \approx 4\times
10^{-14}$~\cite{Eilam:1990zc, Mele:1998ag}, are tiny.  This
extraordinary suppression is caused because, firstly, the
Glashow-Iliopoulos-Maiani (GIM)~\cite{Glashow:1970gm} suppression
prohibits the loop diagram leading contribution for $\tqh$, and
secondly, because the quarks circulating in the $\tqh$ loop amplitude
are those of down type with small mass differences.

On the contrary, in a well motivated extension of the SM, the R-parity
conserving Minimal Supersymmetric Standard Model
(MSSM)~\cite{Nilles:1983ge,Haber:1984rc,Martin:1997ns}, although the
GIM mechanism is still operative in the quark-interactions, it is not,
in general, in the squark interactions.  Eventually, coloured scalars,
the squarks, enter in loops with potentially large mass differences.
The question is then whether these new interactions are able to
enhance $\Br(\tqh)$ up to an observable level at LHC.
Depending on MSSM input parameters, Guasch and
Sola~\cite{Guasch:1999jp} arrived at a maximum prediction ${\Br}(t\to
c\, h) \approx 4 \times 10^{-4}$, while a more recent analysis by Cao
{\it et.al}~\cite{Cao:2007dk,Cao:2006xb}, taking into account
constraints from rare $B$-meson decays, concluded a maximum branching
fraction of up to ${\Br}(t\to c\, h) \approx 6\times 10^{-5}$ (for an
earlier study see also \Ref{DiazCruz:2001gf}).  Finally, not long ago,
a new analysis by the authors of \Ref{Cao:2014udj} concluded a maximum
branching ratio at the level of $O(10^{-6})$ after constraints.

The relevant Lagrangian governing the rare top decays $\tqh$ in the
physical quark basis, after integrating out all heavy degrees of
freedom, is simply,
\begin{equation}
-\mathcal{L} \ \supset \ C_{L}^{(h)} \bar{q}_{R} \, t_{L}\, h \, + \,
 C_{R}^{(h)}\, \bar{q}_{L} \, t_{R}\, h
\ + \ \mathrm{H.c} \;,
\label{lagtuh}
\end{equation}
with dimensionless (Wilson) coefficients $C_{L,R}^{(h)}$.  Note that
in the MSSM $h$ may stand for one of the two CP-even Higgs bosons
denoted as $h,H$, respectively.  Currently LHC sets an upper
bound~\cite{Aad:2014dya,CMStuh}
\begin{equation}
\Br(\tqh)  \leq 0.79\% \;\; \mathrm{(ATLAS)}\;, \qquad
\Br(\tqh)  \leq 0.56\% \;\; \mathrm{(CMS)}\;.  
\end{equation}
This result places rather weak restrictions onto the Wilson
coefficients: $|C_{L}|, |C_{R}| \lesssim 0.1$.  In renormalisable
theories like the MSSM, the coefficients $C_{L}$ and $C_{R}$ would
come from one-loop diagrams involving gluino (or neutralino)-up
squarks, chargino-down squarks and charged Higgs-down quarks.  The
gluino-loop gives the dominant contribution to $\Br(\tqh)$ that
generically is of the order $\alpha_{s}/4\pi \approx 0.01$, which is
by an order of magnitude less than the current bound, but probably
within LHC's projected reach at $\sqrt{s} =14$ TeV with $3000$
fb$^{-1}$~\cite{Agashe:2013hma} (see also note~\cite{ATLAS-tch})
\begin{equation}
\Br(\tqh) \lesssim 2.0\times 10^{-4} \ \Leftrightarrow \ 
|C_{L}|,|C_{R}| \lesssim {\cal O}(1)\times 10^{-2}\;.
\label{LHCproj}
\end{equation}
There are already many phenomenological studies for these decays, a
partial list included in~\cite{AguilarSaavedra:2000aj,
AguilarSaavedra:2004wm,Kao:2011aa, Wang:2012gp, Craig:2012vj,
Chen:2013qta, Atwood:2013ica, Gorbahn:2014sha}.  Very recently
in \cite{Greljo:2014dka,Wu:2014dba}, plausible techniques that
distinguish between $\tuh$ and $t\to c\, h$ have been suggested.  It
is therefore worth looking for MSSM branching fraction predictions
from both rare top decays, $\tuh$ and $\tch$.

The new flavour structure in the MSSM Lagrangian can be parametrized
in terms of supersymmetry soft breaking squark mass matrices
$m_{Q_{L}}, m_{U_{R}}, m_{D_{R}}$ and trilinear holomorphic $A_{U},
A_{D}$ matrices as well as the trilinear non-holomorphic
$A^{\prime}_{U}, A^{\prime}_{D}$ matrices~\cite{Girardello:1981wz,
Hall:1990ac, Borzumati:1999sp,
Hetherington:2001bk}\footnote{Non-holomorphic terms may arise from the
K\"ahler potential non-renormalizable operators like for example $X
X^{\dagger} H_{1}^{\dagger} {Q}_{L} {U}_{R}/M^{3}$ interaction between
MSSM superfields and hidden sector superfield $X$ whose F-term vev,
$\langle F_{X} \rangle$, is responsible for spontaneous SUSY breaking
in the hidden sector.  In contrast, the holomorphic SUSY breaking
terms arise from superpotential non-renormalizable operators like, $X
H_{2} {Q}_{L} {U}_{R}/M$.  If SUSY breaking mediators of mass
$\mathcal{O}(M)$ are very heavy, as for instance in gravity mediated
SUSY breaking scenario where $M=M_{Pl}$, then non-holomorphic terms
($A_{U}^{\prime}$) are negligible compared to the holomorphic ones
($A_{U}$).  However, they could both be of the same order of magnitude
if SUSY breaking happens at low SUSY breaking scales, comparable to
electroweak scale~\cite{Borzumati:1999sp}.}
\begin{eqnarray}
{\cal L}_{\rm MSSM} & \supset & - \widetilde{Q}_{L}^{\dagger}
m_{Q_{L}}^{2} \widetilde{Q}_{L} - \widetilde{U}_{R}^{\dagger}
m_{U_{R}}^{2} \widetilde{U}_{R} - \widetilde{D}_{R}^{\dagger}
m_{D_{R}}^{2} \widetilde{D}_{R} \nonumber \\[1mm]
&+& \left ( H_{2}\: \widetilde{Q}_{L}\: A_{U}\: \widetilde{U}_{R} +
 H_{1} \: \widetilde{Q}_{L}\: A_{D}\: \widetilde{D}_{R}
 + \mathrm{H.c} \right ) \nonumber \\[1mm]
&+& \left ( H_{1}^{\dagger}\: \widetilde{Q}_{L}\:
 A^{\prime}_{U}\: \widetilde{U}_{R} +
 H_{2}^{\dagger} \: \widetilde{Q}_{L}\:
 A^{\prime}_{D}\: \widetilde{D}_{R} + \mathrm{H.c} \right ) \;,
\label{softi}
\end{eqnarray}
where flavour and gauge group indices have been suppressed.  As we
already mentioned, soft breaking terms in~\eqref{softi} may have
non-trivial structure, so that the quark and squark mass matrices
cannot be diagonalized simultaneously in the same flavour basis.
However, a fully generic structure for these matrices is far excluded
by Kaon, charm, and $B$-physics experiments with the
\emph{exception} of the right handed up-squark mass matrix $m_{U_{R}}^{2}$
and the trilinear soft SUSY breaking matrices $A_{U}$ and
$A_{U}^{\prime}$.  For all other matrices $m$ and $A$
in \eqref{softi}, ``flavour'' experiments help to single out four
possible categories:
\begin{enumerate}
\item Minimal Flavour Violation (MFV) assumption~\cite{D'Ambrosio:2002ex,  
Hall:1990ac}: flavour violation arises only from Yukawa matrices
$Y_{U}$ and $Y_{D}$.
\item Almost degenerate $m$'s - their diagonal elements proportional 
to the unit matrix; $A$'s almost diagonal; small off-diagonal terms in
 $m$'s and $A$'s.
\item As in point (2) but $m$'s become hierarchical: 1st and 2nd generation 
are much heavier than the third.
In this case off-diagonal squark mass matrix elements may be of order
one.
\item Alignment: no particular hierarchy among diagonal squark masses,  but
small squark mixing angles, enforced by some symmetries, as required
by experimental constraints.
\end{enumerate}
MFV basically leads to the same suppression pattern for $\tqh$ as in
the SM and therefore no signal observation is expected at
LHC~\cite{Dery:2014kxa}\footnote{This is also due to the fact that no
$\tan\beta$ enhanced top flavour changing decay amplitudes arise in
the MSSM as we will see shortly.}.
We need therefore to depart from MFV.  This is most conveniently done
by considering the dimensionless flavour violating expansion
parameters (commonly called ``mass
insertions'')~\cite{Gabbiani:1996hi, Misiak:1997ei}:
\begin{equation} 
\Delta_{\tilde{X}}^{IJ} \ = \ \frac{(m_{X}^{2})^{IJ}}{\sqrt{(m_{X}^{2})^{II}\, 
(m_{X}^{2})^{JJ}}} \;,
\label{delta}
\end{equation}
which denotes the ratio of flavour-violating squark mass matrix
elements over an average of flavour-conserving squark mass matrix
elements ($\tilde{X}$ can be $\tilde{U}$ or $\tilde{D}$).  It has been
shown in \Ref{Giudice:2008uk} that, for $\Delta F=1$ processes, the
same (in magnitude) $\Delta$-parameter can be used to parametrize
flavour effects in both cases of hierarchical and degenerate squark
masses, although the $\Delta$-parameter may have different meaning in
each case.  We develop a similar technique here in expanding the full
amplitude for $\tqh$ in powers of $\Delta$'s and therefore discussing
cases (2) and (3) in a unified way.

In the fourth case of alignment quark and squark mass matrices are
forced by some approximate flavour symmetry to be diagonalized almost
by the same field rotation.  This means that the { remaining squark}
rotation angles { in the super-CKM basis} are small, but in general,
squark masses are far from degenerate leading to serious constraints
from K-physics.  In any case, having the light Higgs boson mass at
$126$ GeV, one needs pushing the stop mixing angle to the maximal
value.  This situation does not fit naturally to the case of small
mixing angles.  On this ground we will not examine this case.

In fact we shall show below that the LHC projected bound
(\ref{LHCproj}) is impossible to be reached in the general R-parity
conserved MSSM with degenerate or hierarchical squark mass spectrum.
This is partly due to cancellations between self energy and penguin
contributions prohibiting non-decoupling SUSY effects.  As a result,
in the best case scenario, and before constraints, an estimate of the
dominant gluino-squark diagrams results in
\begin{equation}
C_{L,R}^{(h)} \ \approx \ \frac{\alpha_{s}}{4\pi}\, \left
(\frac{m_{t}}{M_{S}} \right)^{2}\, \Delta
\ \lesssim \ 2 \times 10^{-4}\;, \label{appform}
\end{equation}
for degenerate SUSY squark masses $M_{S}$ at 1 TeV scale and $\Delta
={\cal O}(1)$.  Similar cancellations exist in the chargino-squark
loops but now $\alpha_{s}\to \alpha_{2}$ and therefore, following
(\ref{appform}), $C_{L,R}^{(h)}$ are by at least a factor of three
smaller than the gluino contribution.\footnote{In fact chargino
diagrams are far smaller than that because of the down squark
circulation in loop.  The relevant $\Delta_{\tilde{D}}$'s in this case
must be small to respect experimental constraints from low energy
meson experiments.  Similar situation applies to charged Higgs boson
one-loop diagrams.\label{foot:chargino}} Furthermore, as it is obvious
from (\ref{appform}), both our analytical and numerical study
concludes that there are \emph{no} non-decoupling effects whatsoever
for large SUSY mass spectrum, collectively indicated here as $M_{S}$.

To the best of our knowledge, this study deals with four new aspects
of $\Br(\tqh)$ not considered before in the
literature~\cite{Guasch:1999jp, Cao:2007dk, Cao:2006xb,
DiazCruz:2001gf, Cao:2014udj}:
\begin{enumerate}

\item We take into account the effects of Next to Leading order QCD 
corrections due to the SUSY loop induced chromomagnetic dipole
operator and the running of operators from the SUSY scale $M_{S}$ to
the top quark scale (see Section~\ref{calcs}).
 
\item We present analytical details of the cancellations and decoupling
(Section~\ref{cancel}), using a common scheme for both universal and
hierarchical squark mass structures.

\item We investigate the effect on $\Br(\tqh)$ from
non-holomorphic SUSY breaking terms $A_{U}^{\prime}$ [see \eq{softi}]
(Section~\ref{results}).
  
\item  Finally, we have encoded all our calculations into a publicly
available\footnote{\code{} can be downloaded from {\tt
http://www.fuw.edu.pl/susy\_flavor}} \code{}
library~\cite{Rosiek:2010ug, Crivellin:2012jv,Rosiek:2014sia}.
\code{} uses the relevant and most complete up-to-date
constraints from FCNC processes (Section~\ref{constraints}).

\end{enumerate}

\section{Calculation of $\Br(\tqh)$ in MSSM}
\setcounter{equation}{0}
\label{calcs}

The gauge-invariant dimension-6 operator responsible for the decay
$\tqh$ can be, after decoupling of heavy particles, simply written
as\footnote{In full SUSY limit with all Higgs (super)fields present,
the corresponding operator is an F-term and, therefore, holomorphic.
It has the form $O^{(h)} = \left ( H_{1} H_{2} \right) \:
{Q}_{L}^{I}\, u_{R}^{J}\, H_{2} \ + \ {\rm H.c} \;.$ Note that this
operator breaks Peccei-Quin and R-symmetry invariance and therefore
its Wilson coefficient must be proportional to quantities that violate
these symmetries, such as the gluino mass, the trilinear soft SUSY
breaking couplings and the $\mu$-parameter, {\it c.f.} \eq{scaling}.}
\begin{equation}
O^{(h)} = \left ( H^{\dagger} H \right) \: \overline{Q}_{L}^{I}\,
u_{R}^{J}\, \widetilde{H} \ + \ {\rm H.c} \;.
\label{Oh}
\end{equation}
$H$ is the SM Higgs field $SU(2)$ doublet, $\widetilde{H} =
i\sigma_{2} H^{*}$ is its charged conjugate, indices $I$ and $J$
denote quark flavours, $Q_{L}^{I}$ is the left-handed quark $SU(2)$
doublet while $u_{R}^{J}$ is the right-handed up-quark singlet.
$SU(2)$ and $SU(3)$ indices are not shown explicitly.
The effective operator in $O^{(h)}$ is of (pseudo)-scalar form and
affects the renormalizable Yukawa interaction $\overline{Q}_{L}^{I}
u_{R}^{J} \widetilde{H}$.
After electroweak symmetry breaking (EWSB) it results in the effective
Lagrangian~\eqref{lagtuh}.

It was shown recently in \Ref{Zhang:2013xya} that the operator
$O^{(h)}$ mixes through QCD strong interactions with the gluonic
dipole operator that has the form
\begin{equation}
O^{(g)} =
 g_{s} \: \overline{Q}_{L}^{I}\, \sigma^{\mu\nu}\, \lambda^{A} \,
 u_{R}^{J}\, \widetilde{H} \, G_{\mu\nu}^{A} \ + \ {\rm H.c.} \;,
\label{Og}
\end{equation}
where $g_{s} =\sqrt{4 \pi \alpha_{s}}$ is the strong QCD coupling,
$\lambda^{A}$ are the Gell-Mann matrices, while $G_{\mu\nu}^{A}$ is
the $SU(3)$ field strength tensor.
Like the operator $O^{(h)}$, the operator $O^{(g)}$ is also chirality
flipping.
After EWSB it results in the effective Lagrangian term
\begin{equation}
-\mathcal{L} \ \supset \ C_L^{(g)\, IJ}\,
 {\overline{u}}_{R}^{I}\, \sigma_{\mu\nu}\,
\lambda^{A} \,   u_{L}^J \, G^{A\,\mu\nu} \ + \
C_R^{(g)\, IJ}\, {\overline{u}}_{L}^{I}\, \sigma_{\mu\nu}\,
\lambda^{A} \,   u_{R}^J \, G^{A\,\mu\nu} \ +
\ {\rm H.c.} \;.
\label{lagog}
\end{equation}

Having listed all operators needed, we enumerate here our steps in
calculating $\Br(\tqh)$:
\begin{enumerate}
\item Full calculation of the relevant 1-Particle-Irreducible (1PI) 
Feynman diagrams $C_{L,R}^{(h)}$ at scale $M_{S}$, where $M_{S}$ is
the lightest coloured sparticle (squark or gluino) mass.
\item Full calculation of the SUSY induced Wilson coefficient $C_{L,R}^{(g)}$ 
associated with the dipole operator $O^{(g)}$ that mix with strong
(QCD) quantum corrections.
\item Use Renormalization Group Equations (RGEs) with formulae taken 
from~\cite{Zhang:2013xya} to run all operators down to the top mass
scale.
\item Calculate the branching fraction at $m_{t}^{pole}$.
\end{enumerate}
In the next two subsections we append technical details entailed in
these steps.

\subsection{Branching ratio and QCD corrections}
\label{QCD}

In this Section we present the calculation for the decay of the top
quark into a light quark $q=u,c$ and a CP-even Higgs boson $h \equiv
(H~\mathrm{or}~h)$ including NLO QCD corrections.  In the limit
$m_{q}\approx 0$ the tree level decay rate reads:\footnote{Although
straightforward, decays $t\to q\; A$ with $A$ being the CP-odd Higgs
boson are only marginally permitted by recent LHC data and therefore
not considered in this work, {\it c.f., } discussion in
Section~\ref{lma}.}
\begin{equation}
\Gamma_{0}(t\to q \, h) \ = \ \frac{m_{t}}{32 \pi} 
\left ( |C^{(h)}_{L}|^{2} + |C^{(h)}_{R}|^{2} \right ) \:
\left ( 1- \frac{m_{h}^{2}}{m_{t}^{2}} \right )^{2} \;,
\end{equation}
with $C_{L,R}^{(h)}$ defined in \eq{lagtuh}.  At the top-quark mass
scale, $\mu = m_{t}$, the following QCD NLO decay rate is
found~\cite{Zhang:2013xya},
\begin{eqnarray}
\Gamma(t\to q \, h ) \ &=& \ 1.018\, \Gamma_{0}  \nonumber \\
&+& 0.049 \, \frac{m_{t}^{3}}{16 \pi v}\, \left
(1- \frac{m_{h}^{2}}{m_{t}^{2}} \right )^{2} \,
\Re e \left [ C_{R}^{(h)\, *} C_{R}^{(g)} + C_{L}^{(h)\, *} C_{L}^{(g)} \right ]\;,
\label{gammaf}
\end{eqnarray}
with $C_{L(R)}^{(g)}$ defined in~\eq{lagog}.  We use
$\alpha_{s}(m_{t}) = 0.1079$, $m_{t}(m_{t})_{\overline{DR}} = 163.6$
GeV, $m_{t}^{pole} = 172.5$ GeV, $G_{F} = 1/\sqrt{2} v^{2} =
1.1664\times 10^{-5}~\mathrm{GeV}^{-2}$.  In our results we have
neglected terms proportional to $|C_{L,R}^{(g)}|^{2}$ since they are
small for $m_{h} \simeq 126$ GeV.
For the branching fraction $\mathcal{B}(\tqh)$, the
next-to-next-to-leading order top quark width is used, $\Gamma(t\to
bW) = 1.39$ GeV~\cite{Czarnecki:2010gb}.  Furthermore, we assume that
the ``tree level'' decay width $\Gamma(t\to bW)$ is not affected
substantially by SUSY loop contributions.
In this Section, we calculate the Wilson coefficients, $C_{L,R}^{(h)}$
and $C^{(g)}_{L,R}$, at the scale $\mu = M_{\rm S}=m_{\tilde{g}}$, and
use the renormalization group equations~\cite{Zhang:2013xya} to run
them down to the scale $\mu=m_{t}$,
\begin{subequations}
\begin{eqnarray}
C_{L,R}^{(h)}(m_{t}) \ &=& \ C_{L,R}^{(h)}(M_{\rm S})
\left (\frac{\alpha_{s}(M_{\rm S})}{\alpha_{s}(m_{t})} \right )^{-4/b_{3}}  
\nonumber \\[2mm]
&+& \frac{24}{7}\, \frac{m_{t}(m_{t})^{2}}{v}\, C_{L,R}^{(g)} (M_{\rm
S})\, \left[ \left( \frac{\alpha_{s} (M_{\rm S})}{\alpha_{s}
(m_{t})} \right )^{2/(3b_{3})} - \left (\frac{\alpha_{s}(M_{\rm
S})}{\alpha_{s}(m_{t})} \right )^{-4/b_{3}} \right ] , \\[3mm]
C_{L,R}^{(g)}(m_{t}) \ &=& \ C_{L,R}^{(g)}(M_{\rm S}) \,
\left (\frac{\alpha_{s}(M_{\rm S})}{\alpha_{s}(m_{t})} \right )^{2/(3b_{3})}\;,
\label{QCDrun}
\end{eqnarray}
\end{subequations}
where $b_{3} = 11 -2 N_{f}/3$ is the 1-loop gluon $\beta$-function.
In our case $N_{f}=6$, i.e., we assume there are no other coloured
particles below $M_{\rm S}$ except from the six SM quark flavours.
Diagrams that do not involve coloured particles are ``frozen'' at the
$m_{t}$-scale and do not participate in the running of Wilson
coefficients in \eq{QCDrun}.

It turns out that the effect of consistently including NLO QCD
corrections in $\Br(\tqh)$ is about 20\%.  This is primarily due to
the RGE running of $C^{(h)}_{L,R}$ from $M_{\rm S}$ down to the top
quark mass scale, and, secondarily due to finite SUSY corrections in
$C^{(g)}$ present in the decay width (\ref{gammaf}).  The $C^{(h)}$
and $C^{(g)}$ coefficients, although in theory different in their
Dirac and Lorentz structures, are both subject to the same
squark-gluino Feynman diagram contribution.  $C^{(g)}$ has analogous,
and even more persisting, cancellations than $C^{(h)}$, due to the
flavour conserving gluon-squark vertex of the former, and the same
flavour changing insertions i.e., same $\Delta$'s.  As a result, it
turns out that SUSY contributions to $m_{t}\, C^{(g)}$ are at most of
the same order as in $C^{(h)}$ and give an amount of 2-10\% correction
to the decay width.

\subsection{Wilson coefficients : full MSSM corrections}
\label{wilsons}

\begin{figure}[tb]
\centering 
\includegraphics[width=\linewidth]{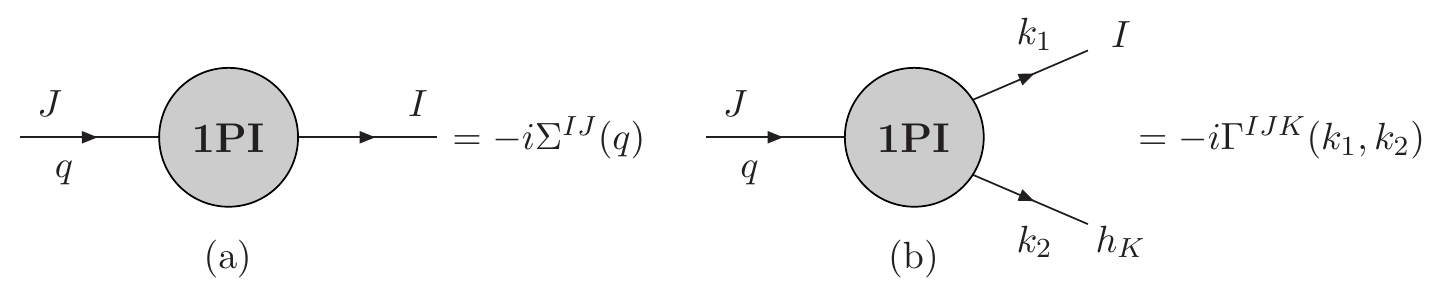}
\caption{\sl (a) Quark self energy one-particle
   irreducible (1PI) diagram corrections.  (b) 1PI penguin
   contribution to $u^{J} \to u^{I} + h_{K}$.
\label{fig:self} }
\end{figure}

Expressions for Wilson-coefficients are more transparent if we write them
in terms of the one particle irreducible (1PI) diagrams for self
energies ($\Sigma$) and penguins ($\Delta F$), as in
Fig.~\ref{fig:self}.  We define:
\begin{eqnarray}
\Sigma^{IJ}(p) =
     \Sigma_{VL}^{IJ}(p^2)\, \slashed{p}\,
    P_L+\Sigma_{VR}^{IJ}(p^2) \,\slashed{p}\,
    P_R+\Sigma_{mL}^{IJ}(p^2) \,P_L+\Sigma_{mR}^{IJ}(p^2) \, P_R\;,
\label{eq:sigdef}
\end{eqnarray}
\begin{eqnarray}
\Gamma^{IJK}(k_1,k_2) = \Delta F_L^{IJK}(k_1,k_2) P_L + 
\Delta F_R^{IJK}(k_1,k_2) P_R \;.
\label{eq:gamdef}
\end{eqnarray}
All $\Sigma$'s and $\Delta F$'s depend on external momenta and
internal masses.  We follow everywhere the Feynman rules, notation and
conventions, from \Refs{Rosiek:1989rs,Rosiek:1995kg}.  For specific
processes, the top-quark is identified with $J=3$ and the
charm-(up-)quark with $I=2 \:(I=1)$, the ``little $h$'' Higgs boson
with $K=2$, the ``big $H$'' with $K=1$, but otherwise we keep the
$I,J$ and $K$ notation as general as possible.

Using standard on-shell renormalization scheme techniques we obtain
for $I\ne J$:
\begin{eqnarray}
C_{L}^{(h)\:IJK} &=& \frac{\eta^{K}}{m_{J}^{2} -
 m_{I}^{2}} \, \biggl \{ m_{I}\: m_{J}^{2} \biggl
 [ \Sigma_{VL}^{IJ}(m_{J}^{2}) - \Sigma_{VL}^{IJ}(m_{I}^{2}) \biggr ]
 + m_{I}^{2}\: m_{J} \biggl [ \Sigma_{VR}^{IJ}(m_{J}^{2})
 - \Sigma_{VR}^{IJ}(m_{I}^{2}) \biggr ]
\nonumber \\[3mm]
&+& m_{I} \: m_{J} \biggr [\Sigma_{mR}^{IJ}(m_{J}^{2})
- \Sigma_{mR}^{IJ}(m_{I}^{2}) \biggr ] + \biggl [ m_{I}^{2
}\: \Sigma_{mL}^{IJ}(m_{J}^{2}) -
m_{J}^{2}\: \Sigma_{mL}^{IJ}(m_{I}^{2}) \biggr ]
\biggr \} \nonumber \\[3mm]
&+& (\Delta F_{L})^{IJK}\;, \label{tchCL}
\end{eqnarray}
and $C_{R} = C_{L} : (L \leftrightarrow R)$.  The parameter $\eta$ is
defined as $\eta^{K} \equiv Z_{R}^{2K}/v_{2}$ with $Z_{R}$ defined
in \eqref{zr}.  The self energy components obey the following
hermicity conditions $\Sigma_{VL(R)}^{JI\star} = \Sigma_{VL(R)}^{IJ}$
and $\Sigma_{mL(R)}^{JI\star} = \Sigma_{mR(L)}^{IJ}$ and explicitly
read in a most compact notation as ($S$=scalar, $F$=fermion):
\begin{subequations}
\begin{eqnarray}
\Sigma_{VL}^{IJ}[p,S,F] &\equiv& \sum_{i,j} V_{uSF,L}^{Iji\,*} 
V_{uSF,L}^{Jji}(B_1+B_0)\,\,[p,m_{S_j},m_{F_i}]\;,\\
%
%
\Sigma_{mL}^{IJ}[p,S,F] &\equiv& \sum_{i,j}m_{F_i} V_{uSF,R}^{Iji\,*}
V_{uSF,L}^{Jji}\,\,B_0\,\,[p,m_{S_j},m_{F_i}]\;, \label{sigmamL}
%
\label{sigmamR}
\end{eqnarray}
\label{sigmaLR}
\end{subequations}
with $L\leftrightarrow R$ for $\Sigma_{VR}^{IJ}$ and
$\Sigma_{mR}^{IJ}$.  Generic vertices $V_{uSF}$ follow the notation of
Appendix A.2 in~\Ref{Dedes:2008iw}.  Explicitly for individual SUSY
particles, their forms, copied from \Ref{Rosiek:1995kg}, are given
in~\ref{App:Vs} for complementarity.  Detailed definitions for
two-point one-loop functions $B_0,B_1$ are given in~\ref{PVfuncs}.

The SUSY-mediated $\tqh$ penguin amplitudes can be classified into two
distinct topologies: (SFS) squark-gluino/neutralino/chargino-squark
and (FSF) chargino/neutralino-squark-chargino/neutralino vertex
diagrams.  They both contribute to the expressions for the
$C_{L,R}^{(h)}$ in \eqref{lagtuh},
\begin{equation}
\Delta F_L  = \Delta F_L^{(SFS)}+\Delta F_L^{(FSF)} \;, 
\end{equation} 
where in a self-explanatory notation
\begin{subequations}
\begin{eqnarray}
\Delta F_L^{(SFS)} &=& \Delta F_L^{(\tilde{D}\chi \tilde{D})}
+ \Delta F_L^{(\tilde{U} \chi^{0}\tilde{U})} + \Delta
F_L^{(\tilde{U}\tilde{g} \tilde{U})} \;, \nonumber\\[2mm]
\Delta F^{(FSF)}_L &=& \Delta F^{(\chi \tilde{D} \chi)}_L + \Delta F^{x(\chi^0 
\tilde{U} \chi^0)}_L \;,
\end{eqnarray}
\end{subequations}
and similar for $\Delta F^{(SFS,FSF)}_R$ with the substitution
$L\leftrightarrow R$.  Each term in the above expressions will be
given by a straightforward substitution in the following compact forms
(again explicit vertices for the generalised $V$'s as well as integral
functions can be found in the appendices.  External momenta follow the
conventions of Fig.~\ref{fig:self}b:
\begin{subequations}
\begin{eqnarray}
\Delta F_{L}^{(SFS)\, IJK} &=& -\sum_{i,j,l} \left\lbrace  
m_{I} \: (V_{HSS}^{Kl\,i}V_{uSF ,L}^{I\,l\,j\,*}V_{uSF
,L}^{J\,i\,j})(C_{12}-C_{11})\right.
\nonumber\\[2mm]
&+&m_{J}\,(V_{HSS}^{Kl\,i}V_{uSF ,R}^{I\,l\,j\,*}V_{uSF
,R}^{J\,i\,j})(C_{11}+C_{0})
\nonumber \\[2mm]
&+&\left.\,m_{F_{j}}(V_{HSS}^{Kl\,i}V_{uSF,R}^{I\,l\,j\,*}
V_{uSF,L}^{J\,i\,j}) \, \,C_{0}\right\rbrace
[k_2,k_1,m_{S_i},m_{S_l},m_{F_j}]\;, \label{CSFS}\\[2mm]
\Delta F_L^{(FSF)\, IJK} &=&-\sum_{i,j,l} \left\lbrace( V_{uSF,R}^{Ij\,l\,*}
V_{FHF,R}^{i\,K\,l}V_{uSF,L}^{Jj\,i})\right.
\,\left(\tilde{C}_0+m_{I}^2\: C_{11}+(m_J^2-m_I^2)C_{12}\right)\nonumber\\[2mm]
&+& m_I m_J(V_{uSF,L}^{Ij\,l\,*} V_{FHF,L}^{i\,K\,l}
V_{uSF,R}^{Jj\,i})\,(C_0+C_{11})\nonumber\\[2mm]
&+&m_{F_l} m_{F_i} (V_{uSF,R}^{Ij\,l\,*} V_{FHF,L}^{i\,K\,l}
V_{uSF,L}^{Jj\,i})\,C_0 \nonumber\\[2mm] 
&+&m_{I} m_{F_i} (V_{uSF,L}^{Ij\,l\,*} V_{FHF,L}^{i\,K\,l}
V_{uSF,L}^{Jj\,i}) \,(C_0+C_{11}-C_{12})\nonumber\\[2mm] 
&+&m_{J} m_{F_i} (V_{uSF,R}^{Ij\,l\,*} V_{FHF,R}^{i\,K\,l}
V_{uSF,R}^{Jj\,i})\,C_{12}\nonumber\\[2mm] 
&+& m_{I} m_{F_l}(V_{uSF,L}^{Ij\,l\,*} V_{FHF,R}^{i\,K\,l}
V_{uSF,L}^{Jj\,i})\,(C_{11}-C_{12}) \label{CFSF}\\[2mm]
&+&m_{J} m_{F_l}(\left.V_{uSF,R}^{Ij\,l\,*} V_{FHF,L}^{i\,K\,l}
V_{uSF,R}^{Jj\,i})\, (C_{0}+C_{12})\right\rbrace
[k_1,k_2,m_{S_j},m_{F_l},m_{F_i}]\nonumber \;.
\end{eqnarray}
\label{DFL}
\end{subequations}
Again, from these expressions one may also derive the corresponding
$\Delta F_R^{(SFS,FSF)}$ by just letting $L\leftrightarrow R$.
Integral functions and vertices are given in \ref{App:Vs}
and \ref{PVfuncs}.  We have checked both analytically and numerically
that the SUSY contributions to $\tqh$ amplitudes $C_{L,R}$ are finite
and renormalization scale invariant.  For our numerical analysis, we
have included all the above full expressions into the \code{} library.

Note that a calculation of the effective Higgs-quark vertices in the
MSSM, however without detailed analysis of their phenomenological
implications for top quark decays, can also be found
in \Refs{Crivellin:2010er,Crivellin:2011jt}.  As discussed there,
effects of resummation from higher order chiral corrections are small
in the up-quark sector and should not change our qualitative
discussion below.  However, as such corrections are implemented
in \code{} library, they can indirectly affect those bounds on the
flavour changing up-squark $\Delta$ parameters which are given by
measurements of processes involving down quarks but sensitive to
up-squarks circulating in loop amplitudes.

\section{$\Br(\tqh)$: cancellations, decoupling and qualitative results}
\setcounter{equation}{0}
\label{cancel}

The formulae given previously, although most general, are quite opaque
and do not allow for, at least qualitative, discussion of possible
cancellation or enhancement effects taking place in coefficients
$C_{L,R}^{(h)}$ of \eq{lagtuh}.  We therefore need to perform some
approximations.

In the limit where $m_{I} = m_{u}(m_{c})\to 0$, the coefficients in
(\ref{tchCL}) can be written simply as ($I=1,2$, $J=3$),
\begin{eqnarray}
C_{L}^{(h)\, IJ} \ &=& \ \Delta F_{L}^{(h)\, IJ} \ - \ \frac{1}{v}\, 
\left (\frac{\cos\alpha}{\sin\beta} \right ) \, \Sigma_{mL}^{IJ}(0)\;, 
\label{CLh} \\[2mm]
C_{L}^{(H)\, IJ} \ &=& \ \Delta F_{L}^{(H)\, IJ} \ - \ \frac{1}{v}\, 
\left (\frac{\sin\alpha}{\sin\beta} \right )\, \Sigma_{mL}^{IJ}(0)\;,
\label{CLH}
\end{eqnarray}
with an obvious substitution $L\leftrightarrow R$ for the coefficients
$C_{R}^{(h)}$, $C_{R}^{(H)}$.  The coefficients that multiply the self
energy 1PI diagrams are not simply proportional to $\tan\beta$ as for
example is the case for the $\bar{b}\, s\, h$ transitions in the MSSM.
In the SM limit, where the CP-odd Higgs mass $M_{A}$ (and therefore
the CP-even Higgs boson mass $M_{H}$) is taken to be much heavier than
$M_{Z}$, we have~\cite{Gunion:1989we}
\begin{equation}
M_{A} \gg M_{Z} \quad : \quad \cos\alpha
\approx \sin\beta \;, \quad \sin\alpha \approx - \cos\beta \;.
\label{SMlim}
\end{equation}
In this case only the decay $\tqh$ (i.e., $K=2$) is relevant and the
amplitude is
\begin{equation}
C_{L}^{(h)\, IJ} \ \overset{\rm SM~limit}{=} \ \Delta F_{L}^{(h)\,
 IJ} \ - \ \frac{1}{v}\, \Sigma_{mL}^{IJ}(0)\;,
\end{equation}
with an analogous formula for $C_{R}^{(h)}$.\footnote{At the moment,
LHC data cannot completely exclude~\cite{Djouadi:2013lra} but
rather \emph{disfavour}~\cite{ATLAS-Hpm} the existence of more than
one Higgs boson lighter than $m_{t}$, with such scenario limited only
to certain ``tuned'' scenarios (see for example~\Ref{Drees:2012fb}).}
In our analytical results for $\tqh$ amplitude below, we shall work
with the general expressions in \eqs{CLh}{CLH} and take the SM-limit
(\ref{SMlim}) when necessary.

Due to the presence of the strong QCD coupling, gluino diagrams are
expected to be dominant.  Their contributions can be deduced easily
from the general expressions in \eqs{sigmamL}{CSFS} and give
\begin{subequations}
\begin{eqnarray}
C_L^{(h_{K})\, IJ} = & -& \frac{2 \alpha_s}{3 \pi} \left
(\frac{m_t}{m_{\tilde{g}}^{2}} \right ) \sum_{i,l=1}^6
V^{Kli}_{HUU} \: Z_U^{(J+3)i\, *} Z_U^{(I+3)l} \left(C_0 +
C_{11} \right) \left[\kappa_2, \kappa_1,r_i,r_l,1 \right] 
\qquad \label{g1} \\[2mm]
&+&\frac{2 \alpha_s}{3 \pi} \left (\frac{1}{m_{\tilde{g}}} \right)
\sum_{i,l=1}^{6} V^{Kli}_{HUU} \:  Z_{U}^{Ji\, *}\,
Z_{U}^{(I+3)l} \,C_{0} \left[\kappa_2,\kappa_1,r_i,r_l,1
\right]\label{g2} \\[2mm]
& + & \frac{2 \alpha_s}{3 \pi} \left(\frac{Z_R^{2K}}{v_2} \right)
m_{\tilde{g}} \sum_{i=1}^{6} Z_{U}^{Ji\, *}
Z_{U}^{(I+3)i}\,B_{0} \left [ 0,r_i,1 \right] \;, \label{g3}
\label{Cgluino}
\end{eqnarray}
\label{glucont}
\end{subequations}
where the first two $(\Delta F_L)$ contributions arise from the gluino
penguin with flipped chirality in the top quark external line and the
gluino internal line respectively, while the last from the self energy
$(\Sigma_{mL})$, gluino diagram.  The symbols in~\eqref{glucont} are
defined in~\ref{App:Vs}.  In particular, the mass dimension one,
Higgs-squark vertex, $V^{Kli}_{HUU}$, can be read explicitly
from \eqref{VHUU} and $Z_U$ is the unitary matrix diagonalizing the
up-squark mass matrix (see~\eqref{eq:massdef}), in the basis where
quarks are diagonal.\footnote{For more details on the exact
definitions of the squark mass and rotation matrices the reader is
referred to \Ref{Rosiek:1995kg}.}
Finally, in \eqref{glucont}, we have changed to a more suggestive form
of Passarino-Veltman (PV) functions with dimensionless parameters,
$\kappa_i\equiv k_i/m_{\tilde{g}}$, $r_i\equiv m_i^2/m_{\tilde{g}}^2$,
by simply factoring out the gluino mass scale (details of the
transformation along with useful properties of the PV functions can be
found in \ref{PVfuncs}).
Note that in the completely universal case (MFV scenario) where
$r_i=const$, the whole gluino contribution \eqref{glucont} vanishes
identically due to unitarity of the $Z_U$-matrices.

It is interesting to check \eqref{glucont} for non-decoupling effects.
As we can see from \eqref{VHUU}, the vertex behaves at most as
$V_{HUU}\sim M_{S}$ and therefore, individually, the last two terms
in \eqref{glucont}, do not decouple separately when all SUSY
parameters are scaled up by the same factor.  However, this
non-decoupling behaviour is not realised because of partial
cancellations between the penguin and self energy contributions given
in \eqref{g2} and \eqref{g3}, respectively.  More specifically,
potentially non-decoupled contributions cancel among each other
leaving behind remnants with $\sim m_{t}^{2}/M_S^2$ as leading
behaviour.
In this section, we will show this behaviour both numerically, in the
full expression, and analytically, up to a certain order in the
relevant expansion. For the following quantitative analysis of
cancellations and the leading order contributions, it is sufficient to
work in the zero external momentum approximation for the penguin and
self energy diagrams.

Before proving the cancellations and estimating the behaviour of
surviving contributions, we open a parenthesis here to present a
useful theorem from matrix algebra.  It says the following:
consider a Hermitian $n \times n$ matrix $A$. The trivial
decomposition $A=A^{0}+\widetilde{A}$, where $A^{0}
= \mathrm{diag}(a_{1}^{0},a_{2}^{0},....,a_{n}^{0})$ contains the
diagonal elements of $A$ and $\widetilde{A}$ contains the non-diagonal
elements of $A$, is always possible.  Let the unitary matrix $U$
diagonalizes the matrix $A$ as $U^{\dagger}AU= D$, where
$D=\mathrm{diag}(d_{1},d_{2}.....,d_{n})$ is a diagonal matrix
containing the eigenvalues of matrix $A$.  If we assume that $f$ is an
arbitrary analytic function, we can write down the following
decomposition of matrix $f(A)$ in powers of $\widetilde{A}$ matrix
elements:
\begin{eqnarray}
[f(A)]_{ij} & = & U_{ik}\, f(d_{k}) \, U_{kj}^{\dagger} \ = \
\delta_{ij}\,f(a_{i}^{0}) \ + \ \left (\frac{f(a_{i}^{0}) - 
f(a_{j}^{0})}{a_{i}^{0}-a_{j}^{0}} \right )\, \widetilde{A}_{ij}
+ \nonumber \\[3mm]
&+& \sum_{\ell =1\,, (\ell \neq i,j)}^{n}
\left (\frac{\frac{f(a_{i}^{0}) - f(a_{\ell}^{0})}{a_{i}^{0} - a_{\ell}^{0}}
-\frac{f(a_{j}^{0}) - f(a_{\ell}^{0})}{a_{j}^{0} -
 a_{\ell}^{0}}}{a_{i}^{0}-a_{j}^{0}}\right )\,
\widetilde{A}_{i \ell}\, \widetilde{A}_{\ell j} +.... \;.  
\label{malgebra}
\end{eqnarray}
In case of degenerate eigenvalues, the ill-defined ratios
in \eqref{malgebra} should be replaced by appropriate derivatives.
%
The first line of \eq{malgebra} has been presented
in~\Ref{Buras:1997ij}\footnote{We would like to thank A.  Romanino for
correspondence on this point.}.
A formal proof of \eq{malgebra} generalised to all orders in powers of
$\widetilde{A}$ and its applications to flavour physics will be given
elsewhere~\cite{DPRST}.

For our purpose here, we only require the implementation
of \eq{malgebra} to the relevant expressions in \eq{glucont}.  The
zero external momentum expansion of self energies and penguins
respectively gives (we use here $\hat{ I}^{IJ}$ as for a Kronecker
``delta'' symbol, to avoid confusion with other notation for
supersymmetric parameters):
\begin{subequations}
\begin{eqnarray}
&& \sum_{i=1}^{6} Z_{U}^{Ji\, *} \, Z_{U}^{I+3,i}\,
B_{0}\left[0,r_{i}\right] \ = \ \hat{\Delta}^{J,I+3} \:C_{0}
\left[0\,;r_{J},r_{I+3},1\right]\nonumber \\
&&~~~ + {\sum_{K=1}^6} \hat{\Delta}^{JK} \hat{\Delta}^{K,I+3}\, \,
D_0\left[0\,;r_J,r_K,r_{I+3},1\right] \nonumber \\
&&~~~ + {\sum_{K,M=1}^6}  \hat{\Delta}^{JK}
\hat{\Delta}^{KM}\,\hat{\Delta}^{M,I+3}\:
 E_0\left[0\,;r_J,r_K,r_M,r_{I+3},1\right]\:+ \ {\cal
 O}\left(\hat{\Delta}^4\right)\:,
\label{b00}
\end{eqnarray}
\begin{eqnarray}
&&\sum_{i,l=1}^{6} Z_{U}^{Ji*} Z_{U}^{I+3,l} Z_{U}^{Ki}
Z_{U}^{Ml*} \,\left\{C_0,C_{11}\right\}\left[0\:;r_i,r_l,1\right]
\  = \ \hat{ I}^{JK}\hat{ I}^{M,I+3} \, \left\{C_0,C_{11}\right\} 
\left[0\:;r_J,r_{I+3},1\right] \nonumber\\
&&~~~~~ + \hat{ I}^{JK}\hat{\Delta}^{M,I+3} \, \left\{D_0,D_{11}\right\}
\left[0\:;r_J,r_M,r_{I+3},1\right] 
+ \hat{ I}^{M,I+3}\hat{\Delta}^{JK}\,\left\{D_0,D_{12} \right\}
\left[0\:;r_J,r_K,r_{I+3},1\right]\:\nonumber\\[2mm]
&&~~~~~ +\hat{\Delta}^{JK}\hat{\Delta}^{M,I+3}\left\{E_0,E_{12}\right\}
\left[0\:;r_J,r_K,r_M,r_{I+3},1\right] \nonumber \\
&&~~~~~ + \hat{ I}^{JK} \sum_{N=1}^6 \hat{\Delta}^{MN}\hat{\Delta}^{N,I+3}\,
\left\{E_0,E_{11}\right\}\left[0\:;r_J,r_M,r_N,r_{I+3},1\right]\nonumber \\
&&~~~~~ + \hat{ I}^{M,I+3} \sum_{N=1}^6 \hat{\Delta}^{JN}
\hat{\Delta}^{NK}\, \left\{E_0,E_{13}\right\}
\left[0\:;r_J,r_N,r_K,r_{I+3},1\right] \
 + \ {\cal O}\left(\hat{\Delta}^3\right)
\label{c00}
\end{eqnarray}
\label{bc}
\end{subequations}
in terms of the diagonal and non-diagonal elements of the
dimensionless squark mass matrix
\begin{equation}
r_K^{2}\equiv\frac{({\cal{M}}_U^2)^{KK}}{m_{\tilde{g}}^2} \: \:,\hspace{0.5cm}
\hat{\Delta}^{KM}\equiv \frac{({\cal{M}}_U^2)^{KM}}{m_{\tilde{g}}^2} 
\: \:\:\,{\scriptstyle{(K\neq M)}}\;,\hspace{0.5cm}\hat{\Delta}^{KK}\equiv 0 \;,
\label{Del}
\end{equation}
respectively. The quartic product of $Z_U$ matrices in \eq{c00}
appears after substituting the explicit form of the $V_{HUU}$ vertex
in \eqref{glucont}.  The ``higher derivative'' PV functions $D,E$ are
defined in ~\ref{PVfuncs}, together with iterative relations
generating them from $B$ and $C$ functions.

An important feature of this ``flavour expansion'' framework, also
noted in \Ref{Giudice:2008uk}, is that it allows for a common
treatment of completely different flavour structures in ${\cal
M}_U^2$.  It may apply with the same efficiency in the ``degenerate''
case where the diagonal elements $r_K$ are considered to be equal and
the mass splitting originates only from $\hat{\Delta}$ or in the
``hierarchical'' case where the mass splitting from $\hat{\Delta}$
adds to a pre-existing hierarchical pattern in $r_K$.

Substituting \eqref{bc} into \eqref{glucont} in zero external momentum
approximation for $\Delta{F}_L$ and using the explicit form of the
vertex $V_{HUU}$ from~\ref{App:Vs}, the aforementioned partial
cancellations between self-energy and penguin can be seen to take
place.  While \eqref{g1} itself has the proper decoupling behaviour,
after adding \eqref{g2} and \eqref{g3} only few terms survive,
remarkably only those with a good decoupling behaviour.  After
cancellations and in the most general case where the non-holomorphic
trilinear couplings $A'_U$ are also present, the scale dependence of
the leading remnants in $C_{L}^{(h)\, IJ}$, will behave as
\begin{align}
\eqref{g2} :& \sim\: A'^{JI}_U\: \frac{\cos(\alpha-\beta)}{\sin{\beta}} 
\:\:\times \:{\cal{O}}\left(\frac{1}{M_S}\right)  
&
\eqref{g1} : \sim \: \delta_{RR}^{JI}\left(
\frac{\cos{\alpha}}{\sin{\beta}}\right) \times
{\cal{O}}\left(\frac{m_t^2}{M_S^2}\right)\nonumber\\[2mm]
&\sim\:(\mu^\star Y^J +
A_U^{'JJ}) \: \delta_{RR}^{JI}\: \frac{\cos(\alpha-\beta)}{\sin{\beta}}
\:\times \:{\cal{O}}\left(\frac{1}{M_S}\right)\:&
\sim\:\sum_{A=1}^3\delta_{RL}^{JA}\delta_{LR}^{AI}
\:\:\left(\frac{\cos{\alpha}}{\sin{\beta}}\right)
\times {\cal{O}}(1)\nonumber\\[2mm]
&\sim\: \delta_{LR}^{JI}\:\left(\frac{\cos{\alpha}}{\sin{\beta}}\right)\:
\times {\cal{O}}\left(\frac{m_t}{M_S}\right) &&\nonumber \\[2mm]
&\sim\: \delta_{LR}^{JJ}\delta_{RR}^{JI}\: 
\left(\frac{\cos{\alpha}}{\sin{\beta}}\right)\:
\times {\cal{O}}\left(\frac{m_t}{M_S}\right) &&\nonumber \\[2mm]
&\sim\sum_{A,B=1}^3\: \delta_{LR}^{JA}\,\delta_{RL}^{AB}
\,\delta_{LR}^{BI}
\left(\frac{\cos{\alpha}}{\sin{\beta}}\right)
\times {\cal{O}}\left(\frac{M_S}{m_t}\right)\;,
&&\nonumber\\
\label{scaling}
\end{align}
where we have expressed our results in terms of the more useful
$3\times 3$ block matrices $\delta$.  These are defined through,
\begin{equation}
\hat{\Delta} \equiv
\left(\begin{array}{cc}
\delta_{LL}&\delta_{LR}\\
\delta_{RL}&\delta_{RR}
\end{array}\right)
 \:\::\:\:\delta_{LR} = 
(\delta_{RL})^\dag\:\:,\:\delta_{LL}^{AA}
= \delta_{RR}^{AA}=0\;, \quad (A\, =\, 1,..3)\;.
\label{eq:Delta}
\end{equation}
The analytic expressions in \eqref{scaling} reveal certain regions in
MSSM parameter space where $\Br(\tqh)$ is enhanced and could be
accessible in the high luminosity LHC data.  We will investigate these
enhanced scenarios in Section~\ref{results}.

The scaling behaviour of the leading contributions presented
in \eqref{scaling} is obtained after considering all SUSY mass
parameters scaling simultaneously as $\sim M_{S}$ and all electroweak
mass parameters as $\sim m_t$ in the full expression for the leading
remnants using \eqref{malgebra}.  Under these assumptions the blocks
of $\hat{\Delta}$ in \eq{eq:Delta} will behave as
\begin{equation}
\delta_{LL}\sim \delta_{RR} \sim{\cal{O}}\left( 1\right)
\:,\:\delta_{LR}\sim {\cal{O}}\left(\frac{m_t}{M_{{S}}}\right)\;, 
\end{equation}
due to the non-uniform scaling of the respective blocks inside ${\cal
M}_U^2$.  Using this observation, it is important to notice that all
leading remnants in \eqref{scaling} scale as ${\cal
O}\left(m_t^2/M_S^2 \right)$, which is straightforward to see for all
contributions besides the $\sim \cos{(\alpha-\beta)}$ terms arising
from \eqref{g2}.  These at first sight seem to exhibit a
non-decoupling behaviour, however, a closer look reveals that the
decoupling is hidden within the quantity
\begin{equation}
\frac{\cos(\alpha-\beta)}{\sin{\beta}} \overset{\textrm{SM-limit}}{\simeq} 
2 \cos{\beta}\cos{(2\beta)}\: \frac{M_Z^2}{M_A^2}\sim {\cal
O}\left({\frac{m_t^2}{M_S^2}}\right)\:\: \label{smlimit}\;,
\end{equation} 
and their contribution can become comparable with all other terms,
obviously subject to the $\tan{\beta}$ value chosen.

The flavour structure of \eqref{scaling} may provide us with useful
guiding information on the leading dependence of $C_L^{(h)\,IJ}$ in
terms of the Lagrangian parameters involved.  For example, for
$\tch$-amplitude [$J=3,I=2$ in \eqref{scaling}], the parameters
directly involved are $A_U'^{32}, \delta_{RR}^{32}$ and $A_U^{32}$
with the last parameter always introduced through the
$\delta_{LR}^{32}$ squark mass matrix element.
At a secondary level, flavour conserving parameters such as $\mu$ or
$\delta_{LR}^{33}\sim A_{t}$ may enter the expressions, however only
as pre-factors of the previous ones.  As a result they modify
substantially the final result of $\Br(\tch)$.  Analogous results hold
for $\Br(\tuh)$, with obvious superscript replacements $2 \rightarrow
1$ into parameters above.
\begin{figure}[t]
\centering
\includegraphics[width=0.48\linewidth]{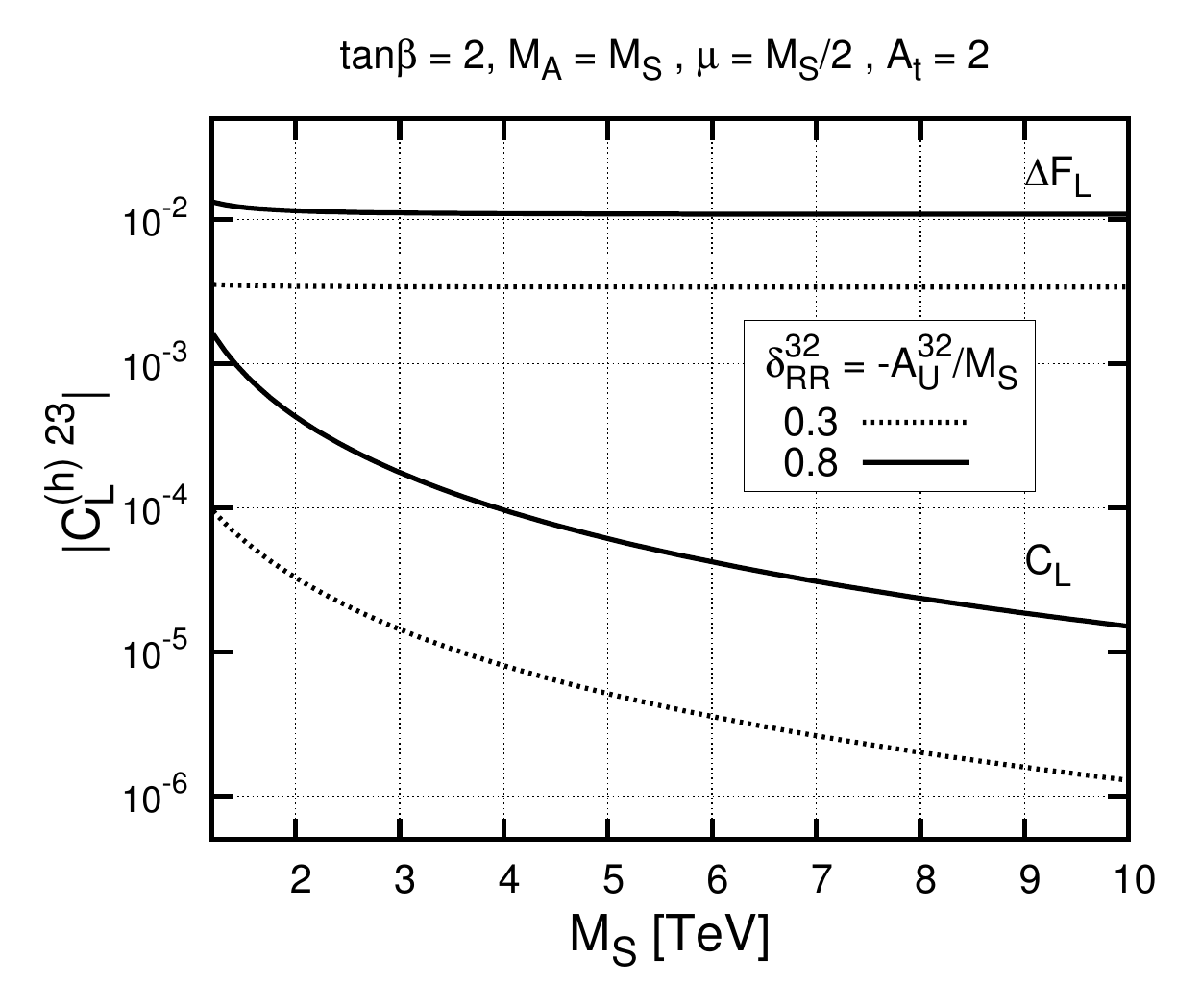}
\includegraphics[width=0.48\linewidth]{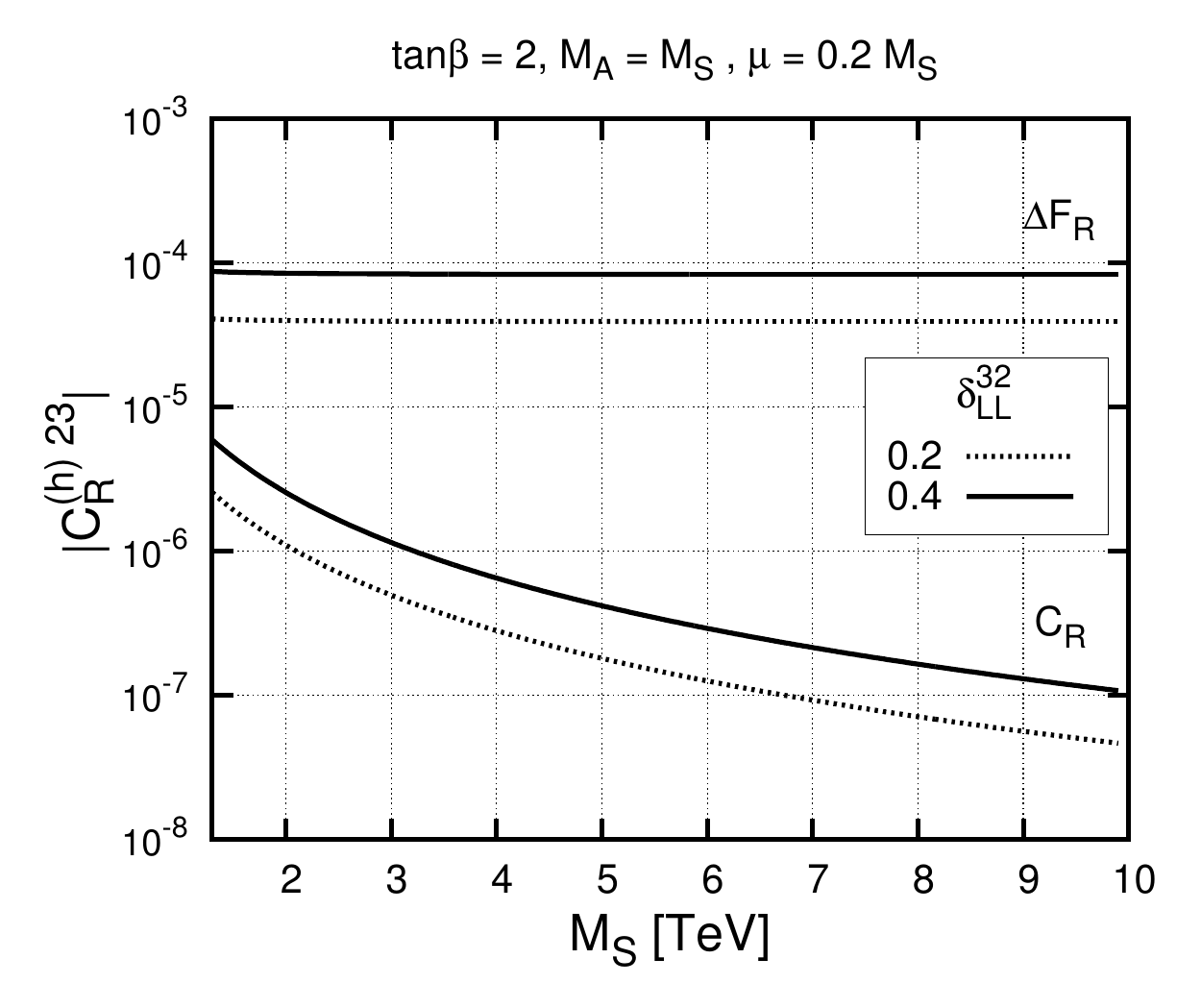}
\caption{\sl (Left) Cancellation and remnants of $|C_{L}^{(h)23}|$  for two  values of the
non-diagonal squark mass parameter $\delta_{RR}^{32}$, assumed here to
 be related to trilinear term as $\delta_{RR}^{32}=-A_U^{32}/M_{S}$,
 in the case of degenerate squark mass spectrum $(r_K^{2}=1)$ and a
 uniform scaling $(M_{A}=m_{\tilde{g}} = M_S)$.  The penguin
 contribution (upper lines) is denoted by $(\Delta F_{L})$.  (Right)
 Similarly, for the Wilson coefficient $|C_{R}^{(h)23}|$ and for two
 values of $\delta_{LL}^{32}$ parameter.  }
\label{Figcanc}
\end{figure}

At this point, it seems instructive to present numerically, in
Fig.~\ref{Figcanc}, the cancellation of self energy and penguin
contributions in $C_L^{(h)\,IJ}$ for a typical choice of the
parameters involved, and for uniform scaling case $M_A=m_{\tilde{g}}=
M_S$ (for the examples illustrated in Fig.~\ref{Figcanc} we ignore
experimental bounds on $\delta$-parameters).  We choose to present
results in $\tch$ amplitude but analogous results hold also for the
$\tuh$ amplitude.
It is clear from Fig.~\ref{Figcanc} (left), where we plot the full
numerical result for the Wilson coefficient $|C_{L}^{(h) 23}|$ with
respect to $M_{S}$, that the non-decoupling behaviour of the penguin
$(\Delta F_L)$ cancels the non-decoupling behaviour of the self-energy
diagrams leaving behind remnants in $|C_L^{(h)\,23}|$ which are
decreasing as $m_{t}^{2}/M_S^2$.  This is scaling behaviour exactly as
our approximate expressions in \eqref{scaling} indicate.

An analogous situation is realised in Fig.~\ref{Figcanc}(right) in the
case of $|C_R^{(h)\,23}|$ for which, due to the aforementioned
$L\leftrightarrow R$ symmetry in the expression for the Wilson
coefficients in \eqref{scaling}, the result primarily depends on
$\delta_{LL}^{32}, A_{U}^{'23\,*}$ and $A_{U}^{23\,*}$.
This clear decoupling behaviour is in qualitative 
agreement with \Ref{Cao:2014udj}, for $M_A=M_S$.

One should note that the terms listed in \eqref{scaling} are leading
or next to leading order contributions in terms of
$\delta$-parameters, obtained in the approximation of vanishing
momenta of the external particles.  Under the uniform scaling of all
SUSY parameters these terms scale as $\sim m_t^2/M_S^2$.  There are
other contributions that scale similarly and can be extracted from the
full amplitude expression \eqref{glucont}.  For example, the first
non-trivial order in the external momentum expansion of the penguin
amplitude $\delta F_{L(R)}(k_1,k_2)$ has the similar flavour structure
and decoupling properties, so it will modify the coefficients of terms
in~\eqref{scaling} but does not change our qualitative discussion.
Other possible terms, e.g.  higher order contributions in the flavour
expansion, are either subleading in $\delta$'s or small due to other
suppression factors, so we do not display them explicitly.  They are
of course included in the numerical analysis presented in next
Sections, as for that we use full unexpanded formulae~\eqref{sigmaLR}
and~\eqref{DFL}.

Finally, similar cancellations of non-decoupling contributions can be
observed numerically (and as we checked also analytically, although
after more complicated calculations) for chargino and neutralino
contributions to the considered $\tqh$ decay amplitude.  Therefore,
they always become smaller than the gluino diagrams, independently of
the soft SUSY breaking parameters scale (see also
footnote~\ref{foot:chargino}).

\section{Constraints from other observables}
\setcounter{equation}{0}
\label{constraints}

As we discussed already in Section~\ref{calcs} we have added our
calculations for $\Br(\tqh)$ into the \code{}
library~\cite{Rosiek:2010ug, Crivellin:2012jv,Rosiek:2014sia}.  For
every input MSSM parameter set, \code{} calculates a number of $B$-,
$K$-, and $D$-meson physics observables.  Comparing them with
experimental bounds~\cite{Beringer:1900zz} allows us to plot
predictions for the $\tqh$ decay rate only for realistic values of the
MSSM parameters.

Most of these observables are related to the processes involving down
quarks and they constrain strongly the flavour structure of
$m_{Q_L}^2$ soft mass matrix, common from both $\tilde{D}$ and
$\tilde{U}$ squarks.  Thus, it is unlikely to have
$\delta_{LL}^{I3}\gtrsim 10$\% and this is impossible to generate
large effects in $\tqh$ decays.  We are therefore going to set
$\delta_{LL}^{i3}$ zero in the numerical results below.
For $\delta_{LR}^{I3}$ and $\delta^{I3}_{RR}$ and at low and moderate
values of $\tan\beta$, potentially important constraints for
$\Br(\tqh)$ arise from the D-meson mass difference, $\Delta M_{D}$.
However, $\Delta M_{D}$ is particularly sensitive to
$\delta_{RR}^{12}$ element, which affects $\Br(\tqh)$ only through
higher powers of $\delta$-insertions than those attributed to the
leading effect in \eqref{scaling}.  Also $\Br(B\to X_{s} \gamma)$ and
$\bar B_{s(d)}-B_{s(d)}$ mixing could be potentially bound to
constraints but they are not significant as contributions from the
right up-squark sector to these processes are suppressed by the powers
of light quark Yukawa couplings.
 
There are of course relevant constraints for parameters important for
$\Br(\tqh)$ emerging from direct, mainly LHC, SUSY
searches~\cite{Aad:2013wta,Beringer:1900zz}.  These are shown in
Table~\ref{tab1}.  A scenario which is particularly interesting for
enhancing $\Br(\tqh)$ is the one with the light stop mostly ``right
handed''.  In this case a lower bound for light stop, together with a
nearly degenerate neutralino, as low as $m_{\tilde{t}_{R}}\approx
200-400$ GeV cannot be excluded in current LHC
data~\cite{Aad:2013wta,Delgado:2012eu,Buckley:2014fqa}.
 
\begin{table}[t]
\begin{center}
\begin{tabular}{cc}
\hline \hline
Quantity &  Current Measurement  \\ \hline
$m_{\tilde{g}}$ & $>$ 1.1~ {\rm TeV}  \\
single light squark  $m_{\tilde{q}}$ & $>$ 500~{\rm GeV}  \\
$m_{\tilde{t}_{L}}$ & $> 600$ ~{\rm GeV}  \\
$m_{\tilde{t}_{R}}$ & $> 200$ ~{\rm GeV}  \\
$m_{h}$ & $(125.9\pm 0.4)$ ~{\rm GeV} \\ \hline
Neutron EDM ($|d_{n}|$) & $<2.9\cdot 10^{-26} ~\mathrm{e~
cm}$~\cite{Baker:2006ts} \\
\hline \end{tabular}
\caption{Experimental bounds used throughout in our numerical analysis.}
\label{tab1} 
\end{center} 
\end{table}

The recent discovery of the Higgs boson mass at
LHC~\cite{Chatrchyan:2012ufa, Aad:2012tfa}, if interpreted as a
``natural'' MSSM light Higgs boson, requires a large, often close to
maximal, trilinear soft breaking coupling $\delta_{LR}^{33} \propto
A_t/M_{S}\approx \sqrt{6}$.  In fact, this helps $\Br(\tqh)$ to be
enhanced as we observe from our qualitative results
in \eqref{scaling}.  We have incorporated in \code{} two-loop
approximate expressions for the CP-even Higgs bosons, based
on \Ref{Heinemeyer:1999be} for contributions from the top/stop sector,
and supplied with results from~\Ref{Haber:1996fp} for contributions
from other sectors.  As stated in~\Ref{Heinemeyer:1999be}, such
approximation should reproduce the full 2-loop result for the Higgs
boson mass with accuracy better than 2 GeV.  Therefore, we allow for a
region $123~\mathrm{GeV} \lesssim m_{h} \lesssim 128~\mathrm{GeV}$,
because of unaccounted theory errors from higher loop corrections.
Note that full 2-loop formula for the MSSM CP-even Higgs boson mass
has not been calculated yet in the fully general flavour violating
case, with large off-diagonal squark mass insertion.  Thus, actual
theoretical error of expressions given in ~\Ref{Heinemeyer:1999be} can
be bigger, affecting the Higgs mass constraints.

\section{Results}
\setcounter{equation}{0}
\label{results}

Our goal here is to find out the maximal outcome on $\Br(\tqh)$ in the
MSSM.  By reading \eqref{scaling} the maximal effect on $\Br(\tqh)$
will be led by the following parameters [FC stands for Flavor Changing
and $A_t(A_t')\equiv A_U^{33}(A_U^{'33})$]:
\begin{align}
\mathrm{Non-FC :} \quad & A_{t}, \quad A'_{t}, \quad m_{\tilde{t}_L}, 
\quad m_{\tilde{t}_R},  \quad\mu, \quad m_{\tilde{g}}\;,  
\quad \tan\beta \;,  \quad M_{A},   \\[2mm]
\mathrm{Holomorphic ~(FC): } \quad & \delta_{LL}^{I3}\;, 
\quad \delta_{RR}^{I3}\;, \quad A_U^{I3} \;, \quad A_U^{3I}\;, 
\label{holfc}\\[2mm]
\mathrm{Non-Holomorphic~(FC) : } \quad &   A_U^{'I3}\;, 
\quad A_U^{'3I}\;.   \label{nonholfc}
\end{align}
Below we present full numerical results mostly for $\Br(\tch)$.  This
is affected by $(I=2)$ parameters in \eqs{holfc}{nonholfc}.  Results
for $\Br(\tuh)$ are exactly the same as one can see from the leading
order expansion~\eqref{scaling} with the obvious replacement $(I=1)$
in the parameters of \eqs{holfc}{nonholfc}.  Constraints from neutron
EDMs are stronger on the latter and as a result we consider mainly
$\Br(\tch)$ in investigating observability at LHC.

As we have remarked earlier, the analytic formulae, \eq{scaling},
allow for the occurrence of enhanced effects in certain regions of
parameter space.  This saves us from time consuming, and often
difficult to understand and interpret, grid-scan plots.  Consequently,
the following possibilities for an enhanced $\Br(\tqh)$ emerge.

\begin{figure}[t]
\centering
\includegraphics[width=0.8\linewidth]{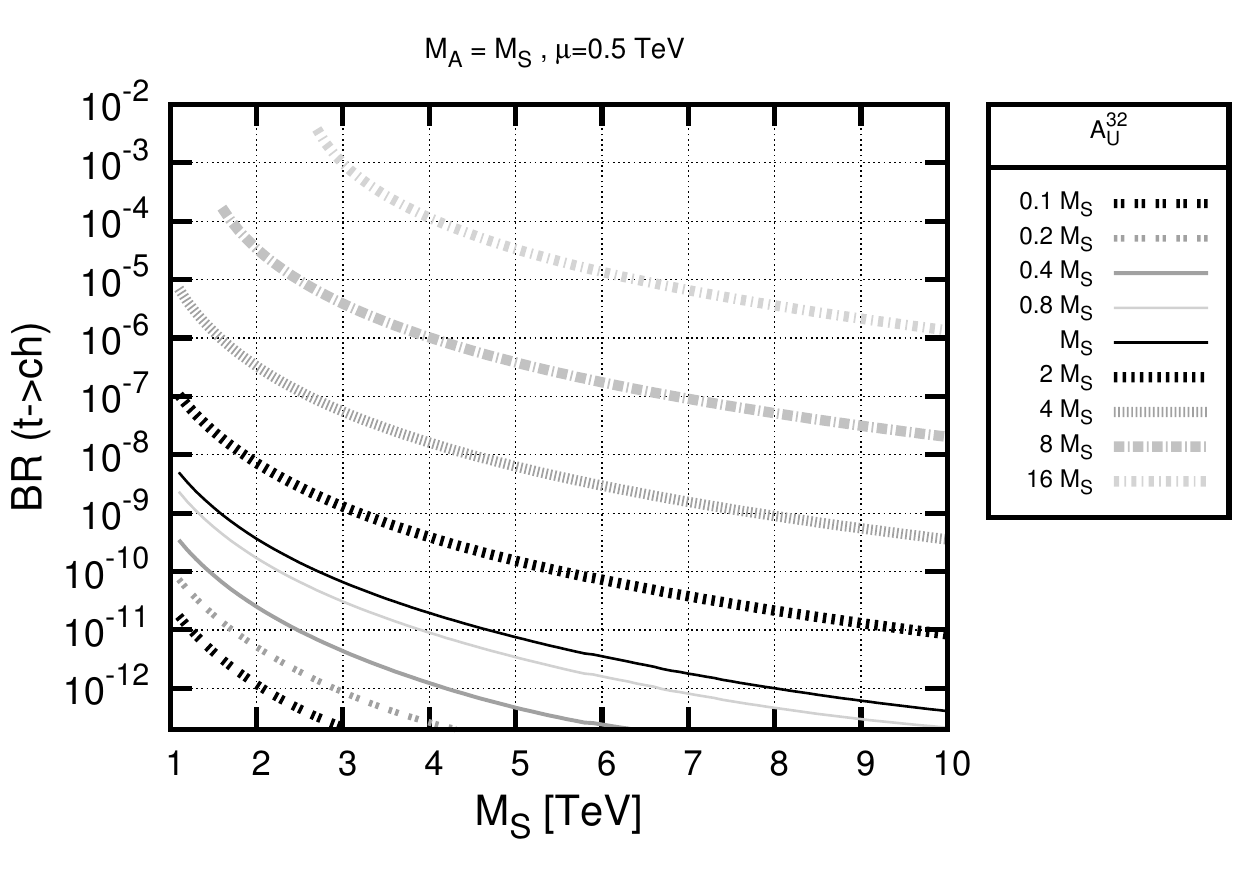}
\caption{\sl Enhancing the $\tch$ decay rate by varying the  $A_U^{32}$ parameter 
for a degenerate spectrum $(r_K=1)$ and a uniform scaling
$(m_{\tilde{g}}=M_A=M_S)$.  $A_{t}\approx 2\, M_{S}$ and
$2 \lesssim \tan\beta \lesssim 4$ are assumed to be consistent with
the measured Higgs boson mass of Table~\ref{tab1}.  The position of
the left edge of each line corresponds to the condition
$m_{\tilde{t}_L}\geq 600$ GeV.  }
\label{Figcanc5}
\end{figure}

\subsection{Enhancement through large $\mathbf{\frac{|A_U^{3I}|}{M_S}}$ and 
$\mathbf{\frac{|A_U^{I3}|}{M_S}}$}

Inspection of~\eqref{scaling} shows that in leading approximation the
expression for $\Br(\tqh)$ contains several terms depending on
up-squark trilinear mixing parameters $A_U^{IJ}$.  The relevant terms
in $C_{L}^{(h)}$ (for $C_{R}^{(h)}$ one needs to exchange chiral
indices $L\leftrightarrow R$) are
\begin{eqnarray}
C_L^{(h)}\;:&&\sim\: \delta_{LR}^{JI}\:\left(
\frac{\cos{\alpha}}{\sin{\beta}}\right)
\: {\cal{O}}\left(\frac{m_t}{M_S}\right)\:,\qquad
\:\:\sim\sum_{A,B=1}^3\: \delta_{LR}^{JA}
\,\delta_{RL}^{AB}\,\delta_{LR}^{BI}
\left(\frac{\cos{\alpha}}{\sin{\beta}}\right)
{\cal{O}}\left(\frac{M_S}{m_t}\right)\;,\nonumber\\
&&\sim\:\sum_{A=1}^3\delta_{RL}^{JA}\delta_{LR}^{AI}
\:\:\left(\frac{\cos{\alpha}}{\sin{\beta}}\right) {\cal{O}}(1)\;.
\label{eq533}
\end{eqnarray}  
Thus, large $A_U^{IJ}$ values can enhance the discussed decay rates.

Such scenario is illustrated in Fig.~\ref{Figcanc5}, where we plot
$\Br(\tch)$ (so that $J=3$ and $I=2$) as a function of $M_{S} =
m_{\tilde{g}} = M_A$ for various values of $A_{U}^{32}/M_S$ and for a
fixed value of $A_t=2M_S$.  In addition, the higgsino mass parameter
is set to $\mu =0.5$ TeV and all other non-diagonal elements of
$\delta$ vanish.  For simplicity in Fig.~\ref{Figcanc5} we vary only
$A_{U}^{32}/M_S$, setting it to several real-positive values, however
as can be seen from analytic formulae, the result for $\Br(\tch)$ is
symmetric under replacement $A_{U}^{32} \leftrightarrow A_U^{23}$ and
depends primarily on the absolute values of both parameters, so we do
not discuss dependence on $A_U^{23}$ separately.

As can be seen from~\eqref{eq533}, the form factor $C_L^{(h)}$ contain
terms with linear dependence and a term with cubic dependence in
$\delta_{LR}^{32}=-\frac{v_2 A_U^{32}}{\sqrt{2}M_S^2}$.  Linear
dependence dominates for $A_U^{32}/M_S\ll 1$ while, more importantly,
cubic dependence dominates for $A_U^{32}/M_S\gg 1$.  As it is obvious
from second line of \eqref{eq533}, the parameter $A_{t}\approx 2M_S$,
required for a $126$ GeV Higgs boson mass, enhances $|C_L|$ and
therefore $\Br(\tch)$, only in parameter regions where linear
dependence dominates, namely for $A_U^{32}/M_S\ll 1$.  In the more
interesting cubic dependence region, where $A_U^{32}/M_S\gg 1$ and the
maximal values of $\Br(\tch)$ are obtained, the branching ratio can
reach LHC attainable values, exceeding estimate \eqref{appform} by two
orders of magnitude, for $A^{32}_{U} \gtrsim 8 M_{S}$ and for a light
$M_S$ value, as can be seen in the left upper corner of
Fig.~\ref{Figcanc5}.  There, the minimum value of $M_S$ is subject to
the condition that the left handed stop squark mass is heavier than
$600$~GeV, as Table~\ref{tab1} indicates.

We should note that our results for $\Br{(\tch)}$ shown in
Fig.~\ref{Figcanc5} do not display any non-decoupling effect.  The
decay rate increases with the $A_U^{32}/M_S$ ratio, but for each fixed
choice of $A_U^{32}/M_S$ it decreases as our analytic formulae
indicate, \ie as $|C_L|^2\sim m_t^4/M_S^4$.  In the most interesting
region $A_U^{32}/M_S\gg 1$, where the cubic dependence in
$\delta_{LR}^{32}$ dominates $C_L$, the branching ratio behaves as
\begin{equation}
\Br{(\tch)}\propto \Big(\frac{A_U^{32}}{M_S}\Big)^6 {\cal{O}} 
\Big(\frac{m_t^4}{M_S^4}\Big)\;.
\end{equation} 
A small deviation from this behaviour can be seen on the left edge of
the upper curves where steeper slopes appear due to
$|\delta_{LR}^{32}|$ closing to unity and higher order corrections
becoming increasingly important.  For large $M_S$, deep in the SM
(decoupling)-limit, although the effect is substantially smaller, the
$\Br(\tch)$ is still enhanced by many orders of magnitude as compared
to the SM prediction.

Another important remark should be done concerning how realistic are
very large values of $|A_U^{32}|/M_S$ (or $|A_U^{23}|/M_S$), required
to enhance the $\Br(\tch)$.  As previously mentioned, they are always
constrained by the condition $|\delta_{LR}^{32(23)}|\lesssim 1$
resulting from the light stop mass bound:
\begin{eqnarray}
|\delta_{LR}^{32}|\sim \frac{v_2}{\sqrt{2} M_S}
\frac{|A_U^{32}|}{M_S}\lesssim  1 \qquad\longrightarrow\qquad
\frac{|A_U^{32}|}{M_S} \lesssim \frac{\sqrt{2}M_S}{v_2}\;.
\label{posit}
\end{eqnarray}
Thus, in principle even very large values of $|A_U^{32}|/M_S$ are
possible assuming sufficiently high SUSY mass scale, e.g.  for
$M_S>1.5$ TeV one can reasonably consider $A_U^{32}\sim 8 M_S$.
However, such large $A_U$ in connection with light stop mass square
can possibly trigger unwanted Charge and Colour Breaking minima
(CCB) \cite{Frere:1983ag, Kounnas:1983td, Gunion:1987qv, Casas:1995pd,
Riotto:1995am, Kusenko:1996jn, Casas:1996de,LeMouel:2001sf,
Park:2010wf, Camargo-Molina:2013sta,Blinov:2013fta,Chowdhury:2013dka}.
For example, allowing for non-vanishing $A_{U}^{32}$ and following the
steps of \Ref{Gunion:1987qv}, and assuming possible vevs in the five
dimensional field space direction, $H_{1}^{0}=0,
H_{2}^{0}=1, \tilde{t}_{L}=0, \tilde{t}_{R}=\tilde{c}_{R}=1$,\footnote{Fields
are normalized to $H_{2}^{0}$ and we take the limit,
$Y_{c}^{2}/Y_{t}^{2} \to 0$.}  we arrive analytically at the following
constraint,
\begin{equation}
|A_{U}^{32}|^{2} \ \le \ Y_{t}^{2} \,(m_{H_{2}}^{2} +
m_{\tilde{t}_{L}}^{2} + m_{\tilde{c}_{R}}^{2} + \mu^{2})
\;,
\label{ccbcon}
\end{equation}
in agreement with \Ref{Casas:1996de}. One can arrive at an even
stronger bound involving both $|A_{t}|$ and $|A_{U}^{32}|$, which is
appreciable because of the Higgs mass constraint, following the field
direction $\tilde{t}_{L} = H_{2}^{0} = 1, H_{1}^{0}=0, \tilde{t}_{R}
= \tilde{c}_{R} = 1/\sqrt{2}$,
\begin{equation}
\bigl ( |A_{t}| + |A_{U}^{32}| \bigr )^{2} \le 4 \; Y_{t}^{2} \; \bigl  [m_{H_{2}}^{2} 
+ m_{\tilde{t}_{L}}^{2} + \frac{1}{2} (m_{\tilde{t}_{R}}^{2} +
m_{\tilde{c}_{R}}^{2} ) \bigr ]^{2} \;,
\label{ccbcon2}
\end{equation} 
in agreement with a similar one found recently
in \Ref{Altmannshofer:2014qha}.
For a common squark and Higgs mass scale, $M_{S}$, the constraint
\eqref{ccbcon} results in $|A_{U}^{32}| \lesssim \sqrt{3} M_{S}$,
which is far more stringent than the positivity physical mass squared
constraint of \eqref{posit}.  For such values of $A_{U}^{32}$, and,
after reading from Fig.~\ref{Figcanc5}, we deduce that
\begin{equation}
\Br(\tch) \ \lesssim  \ 10^{-7}\;.
\end{equation}
This rate is out of any near future LHC expected sensitivity
[see \eqref{LHCproj}].

A detailed analysis of the CCB problem in the general flavour
violating MSSM is beyond the scope of this paper.  Nevertheless, in
most cases the issue is a cosmological one, since sometimes the
inverse transition rate between meta-stable vacua exceeds the lifetime
of the universe.  In this case, the pre-factor of $\mathcal{O}(1)$ in
the RHS of \eq{ccbcon} may be modified, but it is unlikely that it
increases by an order of magnitude or so, necessary to achieve
$\Br(\tch)\sim 10^{-4}$.
This claim is supported by the results of \Ref{Park:2010wf}, where the
bound in \eq{ccbcon} is only marginally relaxed by meta-stability.
For recent accounts on meta-stability of the MSSM vacuum in MFV
scenario, see
\Refs{Camargo-Molina:2013sta,Blinov:2013fta,Chowdhury:2013dka}.\footnote{A 
more robust check for CCB vacua can be studied with the publicly
available code {\tt Vevacious}~\cite{Camargo-Molina:2013qva} which
performs a full numerical check of the potential (meta)stability even
at 1-loop level. A thorough scan of the interesting parameter space
can be however limited by long computer run-time.}

\begin{figure}[t]
\centering
\includegraphics[width=0.8\linewidth]{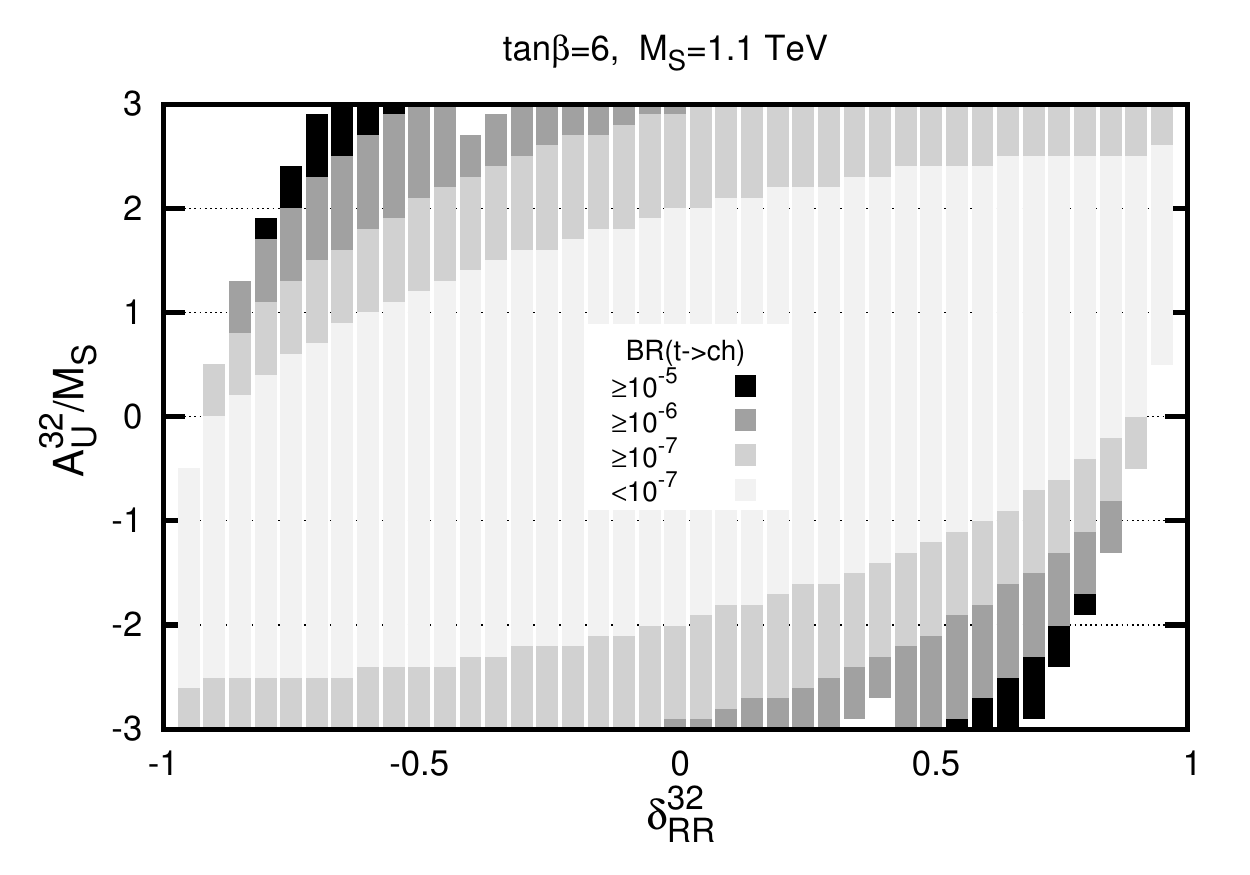} 
\caption{\sl  Contour plot for $\Br(\tch)$ prediction on a 
$\delta_{RR}^{32}$ vs.  $A_{U}^{32}/M_{S}$ plane.  All other FC
parameters are set to zero.  In addition, we take
$(m_{\tilde{g}}=M_A=\mu=M_S)$ and $A_t/M_S=2$ while other values are
shown in the figure.  }  \label{BR4}
\end{figure}

In a more general case both $A_{U}^{32}$ and $\delta_{RR}^{32}$
parameters can be present simultaneously.  In this case the possibly
largest contributions to the $C_L^{(h)}$ form factor, out of all
listed in \eqref{scaling}, are given by terms
\begin{equation}
\sim\: \delta_{LR}^{JJ}\delta_{RR}^{JI}\:\left(\frac{\cos{\alpha}}{\sin{\beta}}
\right)\: \times {\cal{O}}\left(\frac{m_t}{M_S}\right)\;, 
\qquad \sim \: \delta_{LR}^{JI}\,\delta_{RL}^{IJ} \,\delta_{LR}^{JI}
\left(\frac{\cos{\alpha}}{\sin{\beta}}\right)
\times {\cal{O}}\left(\frac{M_S}{m_t}\right)\;.
\end{equation}

In Fig.~\ref{BR4} we plot $\Br(\tch)$ on the $\delta_{RR}^{32}$ and
$A_{U}^{32}/M_{S}$ plane, varying $A_{U}^{32}$ within the region
$|A_U^{32}|/M_{S} \lesssim 3$ in order to avoid potential CCB bounds.
Note that contributions from these two parameters can interfere
constructively (top-left and bottom-right corners of the plot) or
destructively (bottom-left and top-right corners).  However, even in
the most optimistic case, the branching ratio $\Br(\tch)$ \emph{cannot
exceed values of order $\sim 10^{-5}$} which is an order of magnitude
less than the expected sensitivity of LHC.

An analogous effect for $\Br{(\tch)}$ may also arise from the
$C_R^{(h)}$ contribution, namely from the $A_U^{23*}$ and
$\delta_{LL}^{32}$ pair of parameters.  However including such an
effect has little to offer since the enhancement that could be
obtained this way (factor 2 at most) is suppressed due to the
stringent experimental bounds on $\delta_{LL}$.


\begin{figure}[t]
\centering
\includegraphics[width=0.8\linewidth]{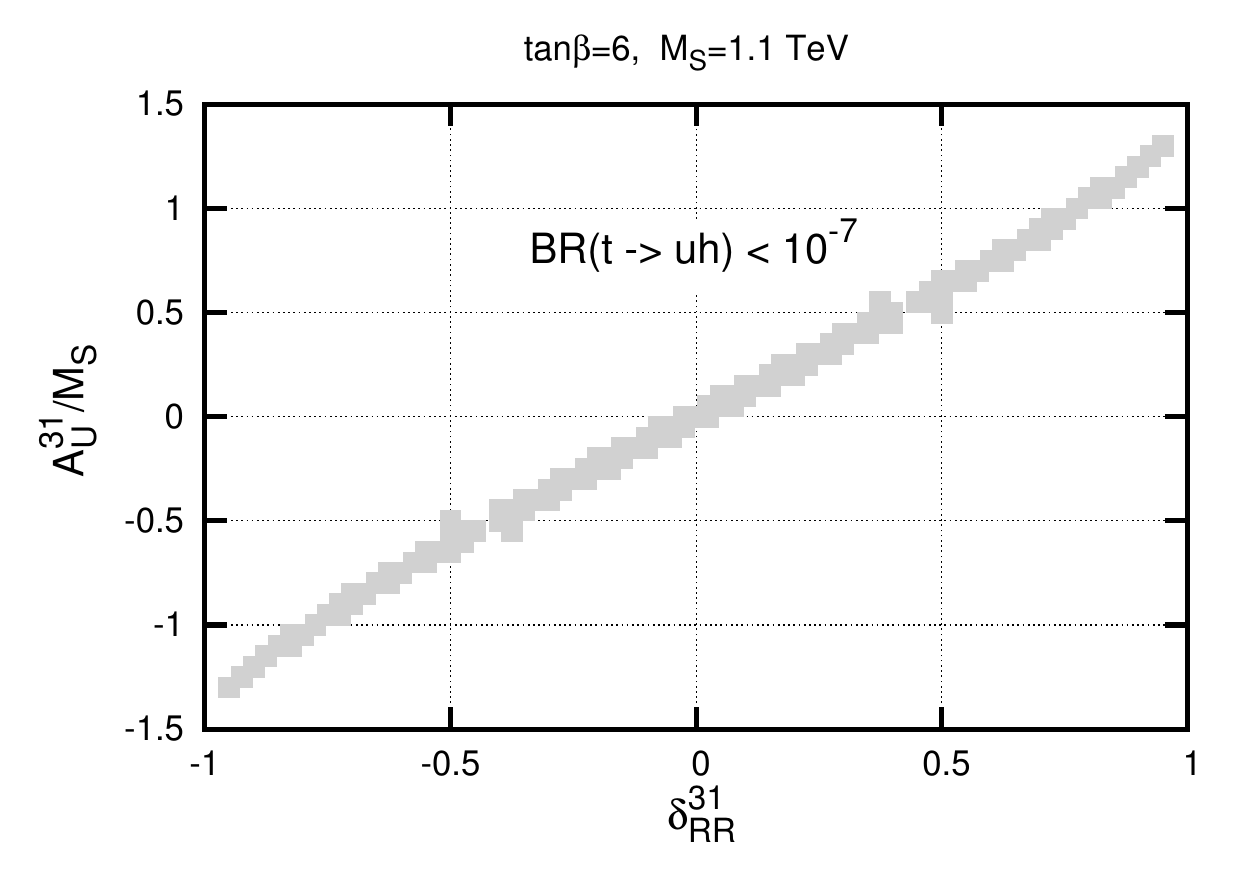} 
\caption{\sl Contour plot for
$\Br(\tuh)$ on a $\delta_{RR}^{31}$ vs.  $A_{U}^{31}/M_{S}$ plane with
all other parameters set as in Fig.~\ref{BR4}.  Due to severe neutron
EDM constraints, the effect is confined to a region where the decay
rate is far beyond the reach of LHC.  }  \label{BR4tuh}
\end{figure}

In Fig.~\ref{BR4tuh} we present results for $\Br(\tuh)$ on a
$\delta_{RR}^{31}$ vs.  $A_{U}^{31}/M_{S}$ plane.  As we have already
discussed, formulae for this decay are exactly the same as for
$\Br(\tch)$, with obvious replacements of indices of flavour violating
parameters.  However, the important difference comes from the fact
that $A_U^{31}$, $ A_U^{13}$ and $\delta_{RR}^{13}$ are highly
constrained by experimental bound on neutron Electric Dipole Moment
(EDM), see e.g.~\cite{Pokorski:1999hz}.  Although we have
assumed \emph{real} parameters throughout this article, this is an
effect that arises from the terms of the higher order in the mass
insertion expansion of the gluino contribution to the down quark
electric and chromoelectric dipole moments.  Such terms are
proportional to $A_U^{31(13)}$ or $\delta_{RR}^{13}$ multiplied by the
CKM matrix elements containing imaginary phase.  Effects of this kind
are usually quite small and unobservable, comparing to experimental
and theoretical accuracy with which most of the rare processes is
known.  However, the bound on neutron EDM is so strong, that it has
visible impact on the acceptable ranges of the real soft parameters.
Approximately, the whole effect results in a strong correlation of the
allowed values of $A_{U}^{31}/M_S$ and $\delta_{RR}^{31}$, such that
their linear combination with ${\cal O}(1)$ coefficients (depending on
up-squark and gluino masses) must vanish with ${\cal O}(10^{-2})$
accuracy, to satisfy the current experimental neutron EDM bound in
Table~\ref{tab1}.  As it is obvious from Fig.~\ref{BR4tuh}, we then
find $\Br(\tuh) \lesssim 10^{-7}$ which is unobservable at LHC.

Based on \eqref{scaling}, one can in principle search how to enhance
$\Br(\tqh)$ other than by previously analysed its cubic dependence on
$A_U^{32}/M_S$.  An analogous effect may also be produced by
increasing $\sim\delta_{RR}^{32}$ together with the unnatural choice
of $|\mu|/M_{S}\gg 1$.  Even so, such a contribution is suppressed by
the condition $\delta_{RR}^{32}<1$ and thus will be typically
subleading, unless $\mu/M_{S}\gg A_U^{32}/M_S$.  Therefore, the
parameter space exploited in Figs.~\ref{Figcanc5} and \ref{BR4} seems
to be the optimal one.
 
Finally, for comparison with the recent literature, we recalculate
results presented in Scan-I of \Ref{Cao:2014udj}. We find numerical
agreement for $\Br(\tch)$ within 10\%. This may be understandable
since we take into account QCD renormalization group running effects
and threshold corrections for Wilson-coefficients neglected
in \Ref{Cao:2014udj} or in other literature quoted in the introduction
section.

\subsection{The light $M_A$ scenario and non-holomorphic dominance} 
\label{lma}

\begin{figure}[t]
\centering 
\includegraphics[width=2.9in]{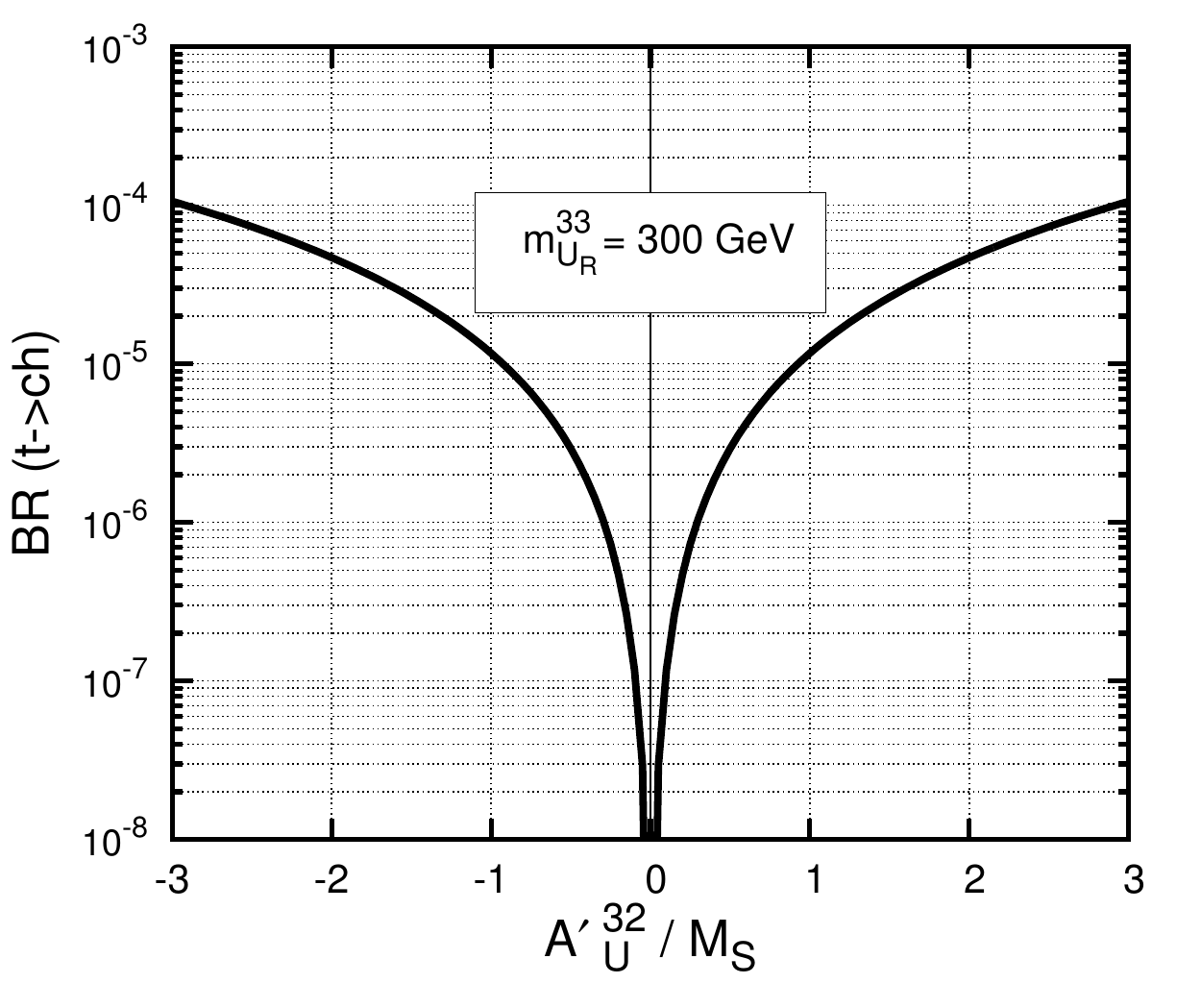} 
\hspace{0.5cm}
\includegraphics[width=2.9in]{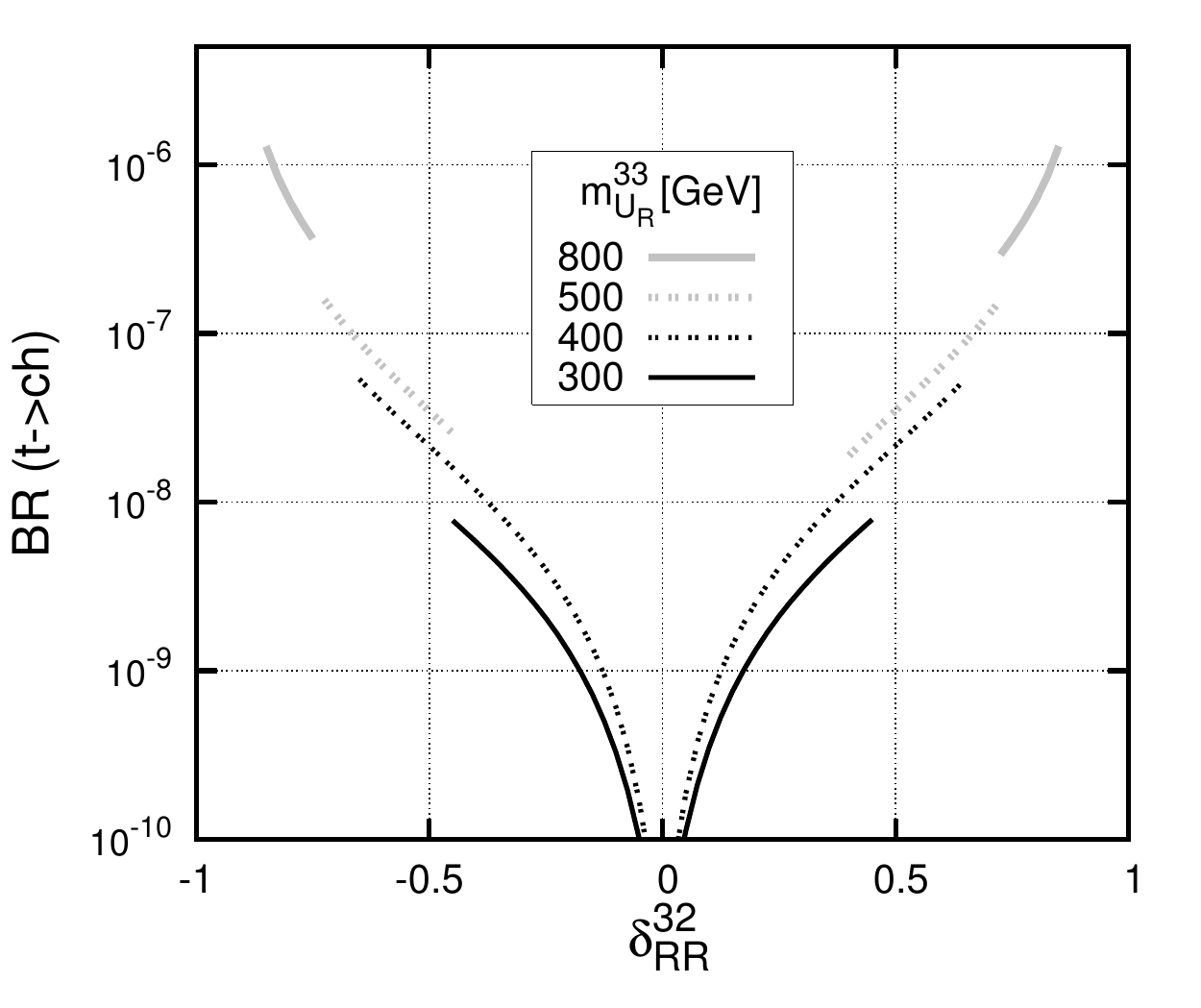} 
\caption{\sl   Branching ratios for the light $M_A=110$ GeV scenario and chosen 
set of MSSM parameters: $\tan\beta=6$, $\mu=250$ GeV, $M_S=1.1$ TeV,
$A_t/M_S=2.7$.  The leading contribution (left panel) originates from
the non-holomorphic coupling ${A'}_U^{32}$.  If ${A'}_U^{32}\approx 0$
the next to leading contribution (right panel) is controlled by
$(\mu^*\delta_{RR}^{32})$.  The allowed $\delta_{RR}^{32}$ range for
each $m_{U_R}^{33}$ value corresponds to the $200 < m_{\tilde{t}_R} <
400 ~\mathrm{GeV}$ constraint of the ``light stop
window"~\cite{Delgado:2012eu}.} \label{BR12}
\end{figure}

The second enhancement scenario requires a light Higgs sector and
significant contribution from the non-holomorphic trilinear soft
couplings, $A_{U}^{\prime}$.  The numerical results are displayed in
Fig.~\ref{BR12}.  We shall attempt here an explanation of the
enhancement based on analytic expansion in~\eqref{scaling}.  We must
warn the reader however that this case scenario is disfavoured by LHC
data and we mostly present it here for complementarity reasons.

This scenario departs from the assumption of uniform scaling for $M_A$
and assume light $M_A\sim M_Z$.  Only terms proportional to
$\cos(\alpha -\beta)$ in~\eqref{g1}, \eqref{g2} will be enhanced [see
also \eq{smlimit}].  To illustrate the size of possible light $M_A$
effects we assume for simplicity vanishing non diagonal squark mass
matrix elements beside $A_U'^{23}$ and $\delta_{RR}^{32}$ in left and
right panels of Fig.~\ref{BR12}, respectively.  In order to make this
point quantitative, we follow the scenario of~\Ref{Drees:2012fb} in
which the heavy Higgs boson is the one seen at LHC with mass around
125.5 GeV and the light one lies in the region $95\lesssim
m_h \lesssim 101\; {\rm GeV}$ where LEP had seen some small excess in
Higgs data.

As we observe from the left panel of Fig.~\ref{BR12}, the
non-holomorphic soft breaking term $A_U'^{23} \approx 3 M_{S}$ may
easily bring $\Br(\tch)$ to the level observable in future LHC
measurements.  This is not true in the right panel of Fig.~\ref{BR12},
where $\delta_{RR}^{32}$ is varied instead.  Here effect is much
smaller due to the constraints $\delta_{RR}^{32} < 1$ (physical squark
masses) and $|\mu|<400\, \mathrm{GeV}$ ($b\rightarrow s \gamma$).  In
that case we obtain $\Br(\tch) \lesssim 10^{-6}$, far off LHC's future
sensitivity.

Also the more promising scenario with enhanced non-holomorphic
contribution in Fig.~\ref{BR12} (left panel) can only be realised in
particular parameter choice.  In such a scenario, the $B\rightarrow
X_{s}\gamma$ constraint imposes $\mu\lesssim 400~\mathrm{GeV}$.  One
of the two charginos, namely the higgsino-like one, with mass
proportional to $\mu$, should be light and cancel the charged Higgs
terms in the respective penguin diagrams.  Thus we choose $\mu$ small
and heavy winos in order to split the chargino masses.
We have also taken a tuned value for trilinear SUSY breaking coupling,
$A_t/M_S=2.7\pm 0.05$, which allows to pass the constraints of
Table~\ref{tab1} and $B\to X_{s}\gamma$, for a large region in $\mu$,
namely $150<|\mu|<350$ GeV.  We could relax the tuning here but only
at the cost of severely restricting the $\mu$ parameter space,
$\mu\simeq (125\sim 150)$.  In any case $|A_t/M_S|\simeq (2\sim 3)$ is
always required in this scenario.  Finally,
following \Ref{Drees:2012fb} we can only vary $\tan\beta$ within the
$\tan\beta\simeq 6\sim 7$ region.

In the light of recent searches for charged Higgs boson produced in
$t\to H^{+}b$ decays and decaying to $\tau$'s~\cite{ATLAS-Hpm} this
scenario seems increasingly unlikely, at least assuming MSSM relations
between the Higgs boson masses.  In principle, there is an open window
(at 1$\sigma$) around $6 \lesssim \tan\beta \lesssim 10$ but only at
very low $m_{H^{\pm}}$ masses less than 110 GeV.  Using the MSSM Higgs
boson mass sum rule $m_{H^{\pm}}^{2} = m_{A}^{2} + m_{W}^{2}$, this
would require very light $M_A \lesssim 75$ GeV far below $M_A=110$ GeV
suggested by the LEP possible excess.

\section{Conclusions}
\setcounter{equation}{0}

In the present article we have studied rare, flavour-changing top
quark decays to light up-quarks $u$ or $c$ and the Higgs boson
$h$, $$t \to u\, h \, \quad \mathrm{or} \quad t\to c\, h \,,$$ in the
framework of MSSM with $R$-parity conservation.
Although the corresponding processes in the framework of the Standard
Model are highly suppressed, mostly due to the GIM mechanism, such a
suppression is not a priori expected in the case of MSSM.

We improve upon existed calculations, most notably from
\Refs{Guasch:1999jp,Cao:2007dk,Cao:2006xb}, by including
next to leading order QCD corrections and RGE running from the SUSY
soft breaking masses down to $m_{t}$.  SUSY finite threshold effects
into $t\to q\, g$ that mixes with $\tqh$, are fully included.  This
set of most up-to-date one-loop corrections to $\tqh$ amplitudes are
then included in publicly available \code{} library, and therefore
combined with MSSM predictions from numerous other flavour physics
observables. In addition to current literature, we study effects
arising from the non-holomorphic soft SUSY breaking terms.  These turn
out to be important for enhancing $\Br(\tqh)$ but in a parameter
region already disfavoured by LHC.

Moreover, we have obtained an analytical expansion of the dominant
gluino amplitude by using a theorem of matrix algebra~\cite{DPRST} and
have arrived at the approximate master formula \eqref{scaling}.  This
formula worked as a guide in order to understand better the
cancellations between various contributions, decoupling effects and
enhancement scenarios in $\tqh$ amplitude.  We conclude that the main
enhancement for $\Br(\tch)$ arises basically from the largeness of the
parameters: $|\delta^{32}_{LR}|\sim |A_{U}^{32}|/M_{S}$ (and/or
$|\delta^{32}_{RL}|$) and $|\delta_{RR}^{32}|$.

Numerical results depicted in Figs.~\ref{Figcanc5} and \ref{BR4} show
that for $|A_{U}^{32}|/M_{S} \gtrsim O(1)$ or $|\delta_{RR}^{32}|\sim
O(1)$, the branching ratio $\Br(\tch) \approx 10^{-5}$ is enhanced
almost by 9 orders of magnitude w.r.t. SM expectation, but
unfortunately it is still below the near future LHC sensitivity
of \eqref{LHCproj}.  This is because of cancellations between leading
order penguin and self energy diagrams, so that decoupling always
takes place.  Only in case where $|A_{U}^{32}| \gtrsim 8 M_{S}$ the
branching ratio is approaching the expected LHC sensitivity. In such a
case however CCB minima are likely to appear as we briefly showed
in \eq{ccbcon} or \eqref{ccbcon2}.

For $\tuh$ on the other hand, although in principle the decay rate is
expected to be of the same order as with $\tch$, the neutron EDM
constraints, induced from the CKM phase, severely suppress the allowed
parameter space into a tuned region in which decay rates are small,
$\Br(\tuh)<10^{-7}$, again far below experimental sensitivity.

We therefore conclude that an MSSM driven $\Br(\tqh)$ is unlikely to
be observed even at high luminosity LHC.  Apart from rather unnatural
corners of the parameter space, the typical MSSM prediction, even for
flavour changing insertions in the up sector of
$\delta_{LR,RR} \sim \mathcal{O}(1)$, is $\Br(\tqh) \approx
10^{-8}-10^{-9}$.  Although small, this is still five to six orders of
magnitude above the SM expectation.
If LHC discovers up-squarks and gluinos it will be vital to develop
techniques that will take us to such small branching ratios for $\tqh$
decay.  If however LHC observes the rare $\tqh$ decays at projected
maximal sensitivity of about $10^{-4}$, their origin must probably lie
in physics other than, or beyond, MSSM with R-parity conservation.

\section*{Acknowledgements}
\setcounter{equation}{0}

AD would like to thank Francesca Borzumati, Sven Heinemeyer and Howie
Haber for useful discussions during SUSY-2014 conference.
We would like to thank Wolfgang Altmannshofer for bringing to our
attention \Refs{Casas:1996de,Park:2010wf,Altmannshofer:2014qha} for
CCB constraints on flavour changing mass insertions.
This research has been co-financed by the European Union (European
Social Fund - ESF) and Greek national funds through the Operational
Program ``Education and Lifelong Learning" of the National Strategic
Reference Framework (NSRF) - Research Funding Program:
THALIS-Investing in the society of knowledge through the European
Social Fund.  J.R.  would like to thank University of Ioannina and
CERN for the hospitality during his stays there.  His work was
supported in part by the Polish National Science Center under the
research grants DEC-2011/01/M/ST2/02466 and DEC-2012/05/B/ST2/02597.

\newpage
\setcounter{section}{0}
\renewcommand{\thesection}{Appendix~\Alph{section}}
\renewcommand{\thesubsection}{\Alph{section}.\arabic{subsection}}
\renewcommand{\theequation}{\Alph{section}.\arabic{equation}}

\section{Explicit expressions for MSSM  vertices}
\setcounter{equation}{0}
\label{App:Vs}

Throughout the paper we follow the notation and conventions
of~\Refs{Rosiek:1989rs, Rosiek:1995kg}, where the definitions of the
Lagrangian parameters, mass matrices and mixing matrices used for
their diagonalization are given for all MSSM sectors.  Here for
completeness we repeat just the explicit expressions for the couplings
needed to calculate the effective $\tqh$ vertex in
Section~\ref{calcs}.  For more details the reader is referred to more
up-to-date \Ref{Rosiek:1995kg}.

The CP-even Higgs boson mass rotation matrix $Z_R$ is defined in terms
of commonly used angle-$\alpha$ as (also as usual $\tan{\beta}
= \frac{v_2}{v_1}$)
\begin{align}
Z_R=\left(\begin{array}{cc}
\cos\alpha & -\sin\alpha \\ 
\sin\alpha & \cos\alpha
\end{array}\right) \,.\label{zr}
\end{align}
Matrices used to mass matrices of supersymmetric particles are
defined, respectively, as:
\begin{eqnarray}
Z_{-}^T M_{C} Z_{+} &=&
diag(m_{\chi_1},m_{\chi_2})\qquad\qquad\qquad \mathrm{chargino}\,,\nonumber\\
Z_N^T M_{N} Z_N &=&
diag(m_{\chi^0_1},\ldots,m_{\chi^0_4})\qquad\qquad \mathrm{~neutralino}\,,\nonumber\\
Z_D^\dagger M_{D}^2 Z_D &=&
diag(m_{D_1}^2,\ldots,m_{D_6}^2)\qquad\qquad \mathrm{down-squarks}\,,\nonumber\\
Z_U^T M_{U}^2 Z_U^* &=&
diag(m_{U_1}^2,\ldots,m_{U_6}^2)\qquad\qquad \mathrm{up-squarks}\,,
\label{eq:massdef}
\end{eqnarray}
where the expressions for $M_{C}, M_{N}, M_{D}^2, M_{U}^2$ can be
found in \Ref{Rosiek:1995kg}.

With the above definitions, relevant tree-level vertices can be
written down as (summation from $1$ to $3$ over all repeating flavour
indices $A,B,\ldots$ is always assumed):
\begin{itemize}
\item Neutral CP-even Higgs-up quark coupling is:
\begin{align}
&V_{uHu}^{IKI}=-\frac{1}{\sqrt{2}}Y_u^I Z_R^{2K} \;,&
\end{align}
where up-quark Yukawa coupling is $Y_u^I=\frac{\sqrt{2}m_u^I}{v_2}$.
\item Couplings relevant for diagram with gluino exchange:
\begin{eqnarray}
V_{HUU}^{Kli}&=&-\frac{e^2}{3c_W^2}(v_1 Z_R^{1K} - v_2 Z_R^{2K})(\hat{
I}^{li} + \frac{3-8s_W^2}{4s_W^2} Z_U^{Al*}Z_U^{Ai}) \nonumber\\
&-&v_2(Y_u^A)^2 Z_R^{2K}(Z_U^{Al\,*}Z_U^{Ai}
+Z_U^{(A+3)l*}Z_U^{(A+3)i} ) \nonumber\\
&+&\frac{1}{\sqrt{2}}Z_R^{2K}(A_u^{AB\,*}Z_U^{Al\,*}Z_U^{(B+3)i}
+A_u^{AB}Z_U^{Ai}Z_U^{(B+3)l\,*})\nonumber\\
&+&\frac{1}{\sqrt{2}}Z_R^{1K}({A'}_u^{AB\,*}Z_U^{Al\,*}Z_U^{(B+3)i}
+{A'}_u^{AB}Z_U^{Ai}Z_U^{(B+3)l\,*})\nonumber\\
&+&\frac{1}{\sqrt{2}}Y_u^AZ_R^{1K}(\mu^*Z_U^{Ai}Z_U^{(A+3)l\,*} +\mu
Z_U^{Al\,*}Z_U^{(A+3)i})\;,
\label{VHUU}\\
\nonumber\\
V_{uU\tilde{g},L}^{Jji}&=&\,\,g_3\sqrt{2}T_{ab}^i(-Z_U^{Jj\,*})\;, \\
V_{uU\tilde{g},R}^{Iji}&=&\,\,g_3\sqrt{2}T_{ab}^i(Z_U^{(I+3)j\,*})\;,
\end{eqnarray}
where the generators $T^i$ are the Gell-Mann of SU(3) with Casimir
invariant normalised to $C^2=\sum_{j} T^j T^j=\frac{4}{3} \,\hat
{\mathbf I}$.

\item Additional couplings necessary for neutralino mediated diagrams are:
\begin{eqnarray}
V_{\chi^0 H\chi^0,L}^{lKi}&=&V_{\chi^0 H\chi^0,R}^{iKl*} = {e \over 2
s_W c_W} \left( (Z_R^{1K} Z_N^{3i} - Z_R^{2K} Z_N^{4i} ) (Z_N^{1l} s_W
- Z_N^{2l} c_W)\right.\nonumber\\
& +& \left.(Z_R^{1K} Z_N^{3l} - Z_R^{2k} Z_N^{4l} ) (Z_N^{1i} s_W -
Z_N^{2i} c_W)\right)\;, \\
V_{uU\chi^0,L}^{Jji}&=& {-e \over \sqrt{2}s_Wc_W} Z^{Ij\star}_{U}
(\frac{1}{3}Z^{1i}_{N} s_W + Z^{2i}_{N} c_W) - Y_u^{I}
Z_{U}^{(I+3)j\star} Z_{N}^{4i} \;,\\
V_{uU\chi^0,R}^{Iji}&=& {2\sqrt{2}e \over 3c_W} Z_{U}^{(I+3)j\star}
Z_{N}^{1i\star} - Y_u^{I} Z_{U}^{Ij\star} Z_{N}^{4i\star}\;.
\end{eqnarray}
\item Couplings relevant for chargino mediated diagrams are:
\begin{eqnarray}
V_{\chi H\chi,L}^{lKi}&=&V_{\chi H\chi,R}^{iKl*} = -\frac{e}{\sqrt{2}
s_W}\left( Z_R^{1K}Z_{-}^{2i}Z_{+}^{1l} +
Z_R^{2K}Z_{-}^{1i}Z_{+}^{2l}\right)\;, \\
V_{uD\chi,L}^{Jji}&=&-(\frac{e}{s_W
}Z_D^{Aj}Z_{-}^{1i}+Y_d^AZ_D^{A+3j}Z_{-}^{2i})\,\,K^{JA\,*}\;, \\
V_{uD\chi,R}^{Iji}&=&\,\,Y_u^I\,\,Z_D^{Aj}Z_{+}^{2i\,*}\,\,K^{IA\,*}\;,
\end{eqnarray}
\begin{eqnarray}
V_{HDD}^{Kli}&=&\frac{e^2}{6c_W^2}(v_1 Z_R^{1K} - v_2 Z_R^{2K})(\hat{I}^{li} 
+ \frac{3-4s_W^2}{2s_W^2}Z_D^{Al*} Z_D^{Ai}) \nonumber\\
&-&v_1(Y_d^A)^2 Z_R^{1K}(Z_D^{Al\,*}Z_D^{Ai}
+Z_D^{(A+3)l*}Z_D^{(A+3)i} ) \nonumber\\
&-&\frac{1}{\sqrt{2}}Z_R^{1K}(A_d^{AB\,*}Z_D^{Ai}Z_D^{(B+3)l\,*}
+A_d^{AB}Z_D^{Al\,*}Z_D^{(B+3)i})\nonumber\\
&+&\frac{1}{\sqrt{2}}Z_R^{2K}({A'}_d^{AB\,*}Z_D^{Ai}Z_D^{(B+3)l\,*}
+{A'}_d^{AB}Z_D^{Al\,*}Z_D^{(B+3)i})\nonumber\\
&-&\frac{1}{\sqrt{2}}Y_d^AZ_R^{2K}(\mu^*Z_D^{Al\,*}Z_D^{(A+3)i} +\mu
Z_D^{Ai}Z_D^{(A+3)l\,*})\;,
\end{eqnarray}
where $K$ is the Cabibbo-Kobayashi-Maskawa matrix and
$Y_d^I=-\frac{\sqrt{2}m_d^I}{v_1}$.
\end{itemize}

One should also note that the conventions used in the paper,
following \Refs{Rosiek:1989rs,Rosiek:1995kg} differ minimally from the
now commonly accepted SLHA2 convention~\cite{Allanach:2008qq} for the
MSSM parameters.  However, translation of the soft breaking parameters
(others do not differ at all) can be done immediately using information from
Table~\ref{tab:slha}.

\begin{table}[htbp]
\begin{center}
\begin{tabular}{|c|c|}
\hline
SLHA2~\cite{Allanach:2008qq} &
Ref.~\cite{Rosiek:1989rs,Rosiek:1995kg}\\
\hline 
& \\[-4mm]
$\hat T_U$, $\hat T_D$, $\hat T_E$ & $-A_u^T$, $+A_d^T$, $+A_l^T$\\
$\hat m_{\tilde Q}^2$, $\hat m_{\tilde L}^2$ & $m_Q^2$, $m_L^2$ \\
$\hat m_{\tilde u}^2$, $\hat m_{\tilde d}^2$, $\hat m_{\tilde l}^2$ &
$(m_U^2)^T$, $(m_D^2)^T$, $(m_E^2)^T$ \\
${\cal M}_{\tilde u}^2$, ${\cal M}_{\tilde d}^2$ & $({\cal M}_U^2)^T$,
$({\cal M}_D^2)^T$ \\[1mm]
\hline
\end{tabular}
\end{center}
\caption{Comparison of~SLHA2~\cite{Allanach:2008qq} and 
Refs.~\cite{Rosiek:1989rs,Rosiek:1995kg} conventions.}
\label{tab:slha}
\end{table}

\section{Passarino-Veltman loop functions}
\setcounter{equation}{0}
\label{PVfuncs}

Our convention for Passarino-Veltman integral functions follows
Axelrod's in \Ref{Axelrod:1982yc}.  For the integrals entering
directly our 1PI-irreducible amplitudes, we have the defining
expressions for 2- and 3-point functions:
\begin{align}
\{B_0,B^\mu\}[k_1,m_1,m_2]&\equiv
\int \frac{d^4p}{(2\pi)^4}
\frac{\{1,p^\mu\}}{(p^2-m_1^2)((p+k_1)^2-m_2^2)}\;,& \label{Bfuncs}\\
\{C_0,C^\mu,\tilde{C}_0\}[k_1,k_2,m_1,m_2,m_3]&\equiv&\nonumber\\
\int \frac{d^4p}{(2\pi)^4}& \frac{\{1,p^\mu,p^2\}}{(p^2-m_1^2) 
((p+k_1)^2-m_2^2) ((p+k_1+k_2)^2-m_3^2)}\;.& \label{Cfuncs}
\end{align}
The expression above can be generalised to the case of general
$n$-point 1-loop functions as:
\begin{align}
&PV_n^{\mu_1\ldots\mu_s}[k_1,\dots,k_{n-1},m_1,\dots,m_{n}]= \nonumber \\[2mm]
&\hskip 30mm \int \frac{d^4p}{(2\pi)^4}\frac{p^{\mu_1}\ldots
p^{\mu_s}}{(p^2-m_1^2)\;\prod_{j=2}^{n} ((p + k_1 + \dots +
k_{j-1})^2-m_{j}^2)}\;, \quad(n\geq 2)\;.
\label{PVn}
\end{align}
Obviously, for $n=2,3$ one obtains the analytic expression for the
$B,C$-functions given in \eqs{Bfuncs}{Cfuncs} (with $\tilde{C}_0=
g_{\mu\nu}PV_3^{\mu\nu}$).  In standard notation higher order
$n=4,5\ldots$ functions are commonly denoted as
$D,E,\ldots$--functions.  Such higher order integrals are absent from
the calculation of $\tqh$ decays at one-loop [\eqs{sigmaLR}{DFL}], but
they unavoidably arise in the flavour expansion approximation
of \eq{bc}.

In practical calculations, it is usually more convenient to replace
the tensorial integral functions by functions transforming as scalars
under the Lorentz group.  For the lowest vectorial functions they are
defined through the relation
\begin{align}
B^\mu&=k_1^\mu \,B_1\;,&&[k_1,m_1,m_2] \label{Bmuscalar}\\
C^\mu&=k_{1}^\mu \,C_{11}+k_2^{\mu}C_{12}\;,&&
[k_1,k_2,m_1,m_2,m_3]\label{Cmuscalar}\\
\ldots\nonumber\\
PV_n^{\mu}& = k_1^\mu\overline{PV}_n^1 + \dots +
k_{n-1}^\mu \overline{PV}_n^{n-1} = \sum_{i=1}^{n-1}
k_i^\mu\overline{PV}_{n}^i \;, && [k_1\dots k_{n-1};m_1\dots
m_{n}]\label{PVbarr}
\end{align}
and similarly for higher tensor functions.  In our notation, all
arguments, common for PV-functions of equal order $n$, are displayed
separately within the respective brackets.

For FC processes, where partial cancellations between topologically
distinct diagrams take place, it is considerably more convenient to
work in a different description of the PV-functions in which all
arguments become dimensionless.  As follows directly from the
definition~\eqref{PVn}, PV loop integrals are homogeneous functions of
their arguments:
\begin{eqnarray}
PV_n^{\mu_1\ldots\mu_s}[k_1,\dots,k_{n-1},m_1,\dots,m_{n}]\; =\;
{M^{4+s-2n}}\;
PV_n^{\mu_1\ldots\mu_s}\left[\frac{k_1}{M}, \dots, \frac{k_{n-1}}{M},
\frac{m_1}{M},\dots,\frac{m_{n}}{M}\right]\;, \nonumber 
\end{eqnarray}
with $M$ being an arbitrary mass scale, usually chosen as a typical
scale for a given loop diagram.

A useful property associates differences of integral functions of a
certain order with integral functions of next order.  For example, as
can be directly verified from the definitions in \eqs{Bfuncs}{Cfuncs},
one has
\begin{equation}
\frac{B_0[k_1,m_1,M_2]-B_0[k_1,m'_1,M_2]}{m_1^2-{m'_1}^2} 
= C_0[0,k_1,m_1,m'_1,M_2]\;.
\label{eq:biter}
\end{equation}
In general case this relation has the following structure:
\begin{eqnarray}
&& {PV_n^X[k_1\dots k_{n-1}; m_1\dots M_{n}] - PV_n^X[k_{1}\dots
k_{n-1}; m'_1\dots M_{n}] \over m_1^2-{m'_1}^2} \nonumber\\
& & \hskip 55mm = PV_{n+1}^X[0,k_{1}\dots k_{n-1}; m_1,m'_1\dots
M_{n}] \;, \label{leq1}\\[4mm]
&& {PV_n^X[\dots k_{j-1}\dots;\dots m_j\dots ] - PV_n^X[\dots
k_{j-1}\dots;\dots m'_j\dots] \over m_j^2-{m'_j}^2} \nonumber\\
&& \hskip 55mm = PV_{n+1}^X[\dots k_{j-1},0\dots ;\dots m_j,m'_j\dots
],\; (j\geq 2)\;,\hspace{1.5cm}\label{lgeq2}
\end{eqnarray}
with $X$ being any set of Lorentz indices of momenta in the numerator
of loop integrand.

For auxiliary scalar functions, defined in \eqst{Bmuscalar}{PVbarr}
this property manifests in a slightly more complicated manner.  That
is, depending on the position of $m_j, m'_j$ within the brackets, the
differences of $\overline{PV}_n^i$ can either produce
$\overline{PV}_{n+1}^i$ or $\overline{PV}_{n+1}^{i+1}$.  For the
lowest order integrals, this property has the suggestive form
\begin{eqnarray}
\frac{B_1[k_1,m_1,M_2] - B_1[k_1,m'_1,M_2]}{m_1^2-{m'_1}^2} = 
C_{12}[0,k_1,m_1,m'_1,M_2] \;, \label{c12aux}\\
\frac{B_1[k_1,M_1,m_2]-B_1[k_1,M_1,m'_2]}{m_2^2-{m'_2}^2}
= C_{11}[k_1,0,M_1,m_2,m'_2]\;.\label{c11aux}
\end{eqnarray} 
For any order of scalar PV functions defined in~\eqref{PVbarr} one has
\begin{align}
&\frac{\overline{PV}_n^i[\dots k_{j-1}\dots;\dots m_j\dots
]-\overline{PV}_n^i [\dots k_{j-1} \dots; \dots
m'_j\dots]}{m_j^2-{m'}_j^2}\nonumber \\[2mm]
&\hspace{65mm}\overset{(j > i)}{=}\overline{PV}_{n+1}^i[\dots
k_{j-1},0\dots ;\dots m_j,m'_j\dots ]\; \nonumber\\
&\hspace{65mm}\overset{(j \leq i)}{=} \overline{PV}_{n+1}^{i+1}[\dots
k_{j-1},0\dots ;\dots m_j,m'_j\dots ]\;.
\label{eq:pviter}
\end{align}

Formulae \eqref{eq:biter}-\eqref{eq:pviter} are particularly useful
because their RHS's are explicitly regular in the limit of degenerate
masses (as all functions defined by 1-loop integrals). Thus, they
allow to generalise \eqref{malgebra} to the case of mass matrices with
degenerated diagonal elements.

\newpage 

\bibliography{biblio-t2uh}{}
\bibliographystyle{JHEP}

\end{document}